\documentclass{JHEP3} \keywords{Beyond Standard Model, NLO Computations} \preprint{Cavendish--HEP--08/17\\
MCnet/09/01}

\usepackage{pstricks}
\usepackage{axodraw}
\usepackage{cite}
\usepackage{amssymb}
\usepackage{amsmath}
\usepackage{subfigure}
\usepackage{color}
\usepackage{slashed}
\let\normalcolor\relax
\usepackage{inputenc}
\usepackage[dvips]{graphicx}
\usepackage{epsfig}
\usepackage{xspace}
\inputencoding{latin1}
\usepackage{epstopdf}

\def\gev{~{\rm GeV}}
\def\tev{~{\rm TeV}}

\def\lsim{\mathrel{\raise.3ex\hbox{$<$\kern-.75em\lower1ex\hbox{$\sim$}}}}
\def\gsim{\mathrel{\raise.3ex\hbox{$>$\kern-.75em\lower1ex\hbox{$\sim$}}}}
\def\ifmath#1{\relax\ifmmode #1\else $#1$\fi}

\title{NLO production of $\boldsymbol{W'}$ bosons at hadron colliders using the MC@NLO and POWHEG methods}

\author{A.~Papaefstathiou and O.~Latunde-Dada\\
Cavendish Laboratory, University of Cambridge, \\
Cambridge, UK\\\\
Email: \email{andreas@hep.phy.cam.ac.uk, seyi@hep.phy.cam.ac.uk}
}

\abstract{We present a next-to-leading order (NLO) treatment of the production of a new charged heavy vector boson, generically called $W'$, at hadron colliders via the Drell-Yan process. We fully consider the interference effects with the Standard Model $W$ boson and allow for arbitrary chiral couplings to quarks and leptons. We present results at both leading order (LO) and NLO in QCD using the \texttt{MC@NLO/Herwig++} and \texttt{POWHEG} methods. We derive theoretical observation curves on the mass-width plane for both the LO and NLO cases at different collider luminosities. The event generator used, \texttt{Wpnlo}, is fully customisable and publicly available. }

\begin{document}

\section{Introduction}\label{sec:intro}
There exists a proliferation of theories which contain new heavy, neutral or charged, gauge bosons referred to as $Z'$ and $W'$ respectively. Both the $Z'$ and $W'$ have been studied extensively and recent reviews can be found in~\cite{pdgreview} and~\cite{zreview}. The present study focuses on $W'$ bosons. The new charged vector bosons may or may not have similar properties to the SM bosons, depending on the theory that predicts them\cite{pdgreview}. In particular they may have right-handed instead of left-handed couplings, may couple to new fermions, or may even be fermiophobic. Popular models which predict new charged vector bosons are based on extensions of the electroweak gauge group, $SU(2)\times U(1)$, for example to the gauge group $SU(2)_1 \times SU(2)_2\times U(1)$ \cite{Mohapatra:1974hk}, or groups that contain the electroweak symmetry, such as $SU(3) \times U(1)$ or $SU(4) \times U(1)$ \cite{Pisano:1991ee}. Several models with extra dimensions contain $W'$ bosons as Kaluza-Klein excitations in the bulk. Examples of these models include the Randall-Sundrum model with bulk gauge fields \cite{Randall:1999ee} and Universal Extra Dimensions \cite{Appelquist:2000nn, Cheng:2002ab}. Theories which break the electroweak sector dynamically may also contain the $W'$ as a composite particle~\cite{tech1, tech2}.  

Current Monte Carlo simulations of Drell-Yan $W'$ production at hadron colliders rely mainly on leading order QCD matrix elements and parton showers.  There currently exists no treatment of next-to-leading (NLO) QCD effects which simultaneously includes the full interference effects for the $W'$. In the present paper, we present the results of the event generator package \texttt{Wpnlo}~\cite{webpage} which improves the treatment of Drell-Yan production of heavy charged gauge bosons. We consider the interference effects with the Standard Model $W$, which have been shown to provide valuable information~\cite{Rizzo:2007xs}, but have not yet been considered in experimental searches. We use the `Monte Carlo at Next-to-leading Order' method~\cite{mcnlo, Frixione:2008ym} for the Herwig++ general purpose event generator~\cite{Bahr:2008pv} (\texttt{MC@NLO/Herwig++}) and the `Positive Weighted Hardest Emission Generation' method (\texttt{POWHEG})~\cite{powheg, Nason:2004rx} to match the NLO QCD calculation to the parton shower, producing fully exclusive events. Note that a similar implementation of the $Z'$ exists for the NLO \texttt{MC@NLO} event generator, which matches the complete NLO matrix elements with the parton shower and cluster hadronization model of the Fortran \texttt{HERWIG} event generator~\cite{zmcnlo}.

The paper is organised as follows: In Section~\ref{sec:lo} the leading order reference model is presented in detail, including the relevant assumptions, and the $W-W'$ interference effects are studied. In Section~\ref{sec:nlo} the \texttt{MC@NLO/Herwig++} and \texttt{POWHEG} NLO implementations of the $W'$ reference model are discussed. Section~\ref{sec:results} presents a sample of distributions obtained using the \texttt{Wpnlo} event generator package. In Section~\ref{sec:limits} we present a theoretical discussion of the extraction of mass-width observation limits for the $W'$ at LO and NLO. We present our conclusions in Section~\ref{sec:conclusion}. The appendix contains the description of a theoretical analysis for discriminating between models.
\section{\boldmath $W'$ at leading order} \label{sec:lo}  
\subsection{The reference model}\label{sec:refmodel}
\begin{figure}[htb]
  \begin{center}
  \input{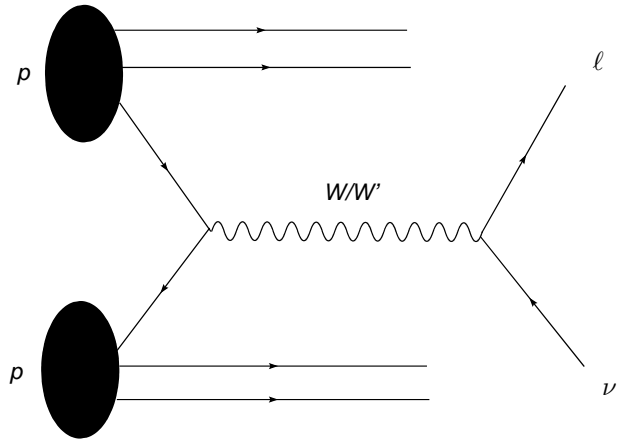}
  \end{center}
\caption{Feynman diagram for $pp \rightarrow W/W' \rightarrow \ell \nu X$.}
\label{fig:ppWlnu}
\end{figure}
The $W'$ reference model is based on the one which originally appeared in\cite{Altarelli:1989ff}. In the model described therein, sometimes referred to as the Sequential Standard Model, the $W'$ couplings to fermions are directly transcribed from the SM $W$, i.e. it is a heavy copy of the SM $W$. In the present treatment we allow both right- and left-handed couplings, corresponding to $(1\pm \gamma_5)$ respectively, as well as arbitrary mixtures of the two. In the case of right-handed couplings, we assume that the right-handed neutrinos are light compared to the $W'$, but not light enough for the $Z$ boson to decay into them. The $W'$ and $W$ couplings to fermions are given by (for $i = W, W'$ ):
\begin{equation}\label{eq:coupling}
\left(\frac{G_FM_W^2}{\sqrt{2}}\right)^{1/2}V_{ff'}C_i^{\ell,q}\bar{f}\gamma_{\mu}(k_i - h_i \gamma_5)f'W^{\mu}_i + h.c.
\end{equation}
where $G_F$ is the Fermi coupling constant, $M_W$ is the SM $W$ mass, $C^{\ell,q}_i$ are the coupling strengths of boson $i$ to leptons and quarks respectively, $W^{\mu}$ is the massive boson polarization vector, $f$ and $f'$ are the Dirac spinors for the fermions and $V_{ff'}$ is the unit matrix when $ff'$ are leptons and the CKM matrix when $ff'$ are quarks. The $k_i$ and $h_i$ represent the structure of the vector-axial vector (V-A) coupling of the bosons, where for the case $i=W$ we have $k_W=h_W = 1$, i.e. purely left-handed coupling. Using the above coupling to fermions, it can be shown that the differential cross-section $pp \rightarrow W/W' \rightarrow \ell \nu X$ (figure~\ref{fig:ppWlnu}) for ($W^+$, $W'^+$), is given by: 
\begin{equation}\label{eq:xsection}
\frac{\mathrm{d} \sigma}{ \mathrm{d} \tau \mathrm{d} y \mathrm{d} z} = \frac{G_F^2M_W^4}{192 \pi} \sum_{qq'}|V_{qq'}|^2 [S G^{+}_{qq'}(1+z^2) + 2AG^{-}_{qq'}z]
\end{equation}
where $z=\cos \theta$ is defined as the scattering angle between the u-type quark and the outgoing neutrino (both being fermions) in the centre of mass (COM) frame, $y$ is the rapidity of the intermediate boson, $\tau = \hat{s} / s$ is the ratio of the squares of the quark COM energy to the proton COM energy. $S=S(\hat{s})$ and $A=A(\hat{s})$ are functions of the quark COM energy.
\begin{equation}\label{eq:Sterm}
S = \sum_{i,j=W,W'}S_{i,j} = \sum_{i,j=W,W'}P_{ij} (C_iC_j)^\ell (C_iC_j)^q (k_ik_j + h_ih_j)^2
\end{equation}
\begin{equation}\label{eq:Aterm}
A = \sum_{i,j=W,W'}A_{i,j} = \sum_{ij=W,W'}P_{ij} (C_iC_j)^\ell (C_iC_j)^q (k_jh_i + h_jk_i)^2
\end{equation}
where 
\begin{equation}\label{eq:pterm}
P_{ij} = \hat{s} \frac{ (\hat{s} - M_i^2) (\hat{s} - M_j^2) + \Gamma_i \Gamma_j M_i M_j } { [(\hat{s} - M_i^2)^2 + \Gamma_i^2 M_i^2][i\rightarrow j]}
\end{equation}
where $M_i$, $\Gamma_i$ are the mass and width of boson $i$ respectively. The functions $G^{\pm}_{qq'}$ which appear in the differential cross-section are even or odd products of parton density functions for the relevant hadrons given by:
\begin{equation}\label{eq:Gpm}
G^{\pm}_{qq'} = \left[q_A(x_a, \hat{s})q'_B(x_b, \hat{s}) \pm q_B(x_b, \hat{s})q'_A(x_a, \hat{s})\right]
\end{equation} 
where $q_{A,B}(x,\hat{s})$ is the parton density function for a quark $q$ carrying hadron momentum fraction $x$ in a collision in which the quark pair COM energy is $\hat{s}$. The $A,B$ indices represent the `left' (travelling in the positive $z$-direction) or `right' (travelling in the negative $z$-direction) hadrons respectively. This definition allows for easy modification of the $pp \rightarrow W/W' \rightarrow \ell \nu X$ cross-section to the $p\bar{p} \rightarrow W/W' \rightarrow \ell \nu X$, by changing the PDFs accordingly. Analogous expressions can also be written in the case of the ($W^{-}$,$W'^{-}$) by appropriately modifying the functions $G^{\pm}_{qq'}$ and taking $z\rightarrow-z$.
The width can be taken to be a free parameter in the reference model: the couplings of the $W'$ to other gauge bosons or the Higgs boson are model-dependent~\footnote{An exception is the photon, for which the coupling is fixed by gauge invariance.}. We shall assume here for illustration that the fermionic decay width scales with the mass as $\Gamma_{W'\rightarrow ff'} = (4\Gamma_{W}/ 3M_{W}) M_{W'}$ (provided that $M_{W'} \gg M_{t}$, the mass of the top quark) and that the tri-boson $W'WZ$ vertex is suppressed by a small mixing angle and hence can be neglected in the analysis.
\subsection[$W-W'$ Interference]{\boldmath $W-W'$ interference}
The narrow width approximation (NWA) is often used when discussing the production of new vector bosons. This approximation is usually claimed to be valid up to $\mathcal{O}(\Gamma_{W'}/M_{W'})$ corrections. But $W$-$W'$ interference effects can become important in certain regions even as the width $\Gamma_{W'}\rightarrow 0$, see for example~\cite{Rizzo:2007xs}, and as we also show below. Use of the NWA may thus lead to invalid conclusions, as pointed out in~\cite{Berdine:2007uv}.

\begin{figure}[htb]
  \begin{center}
  \input{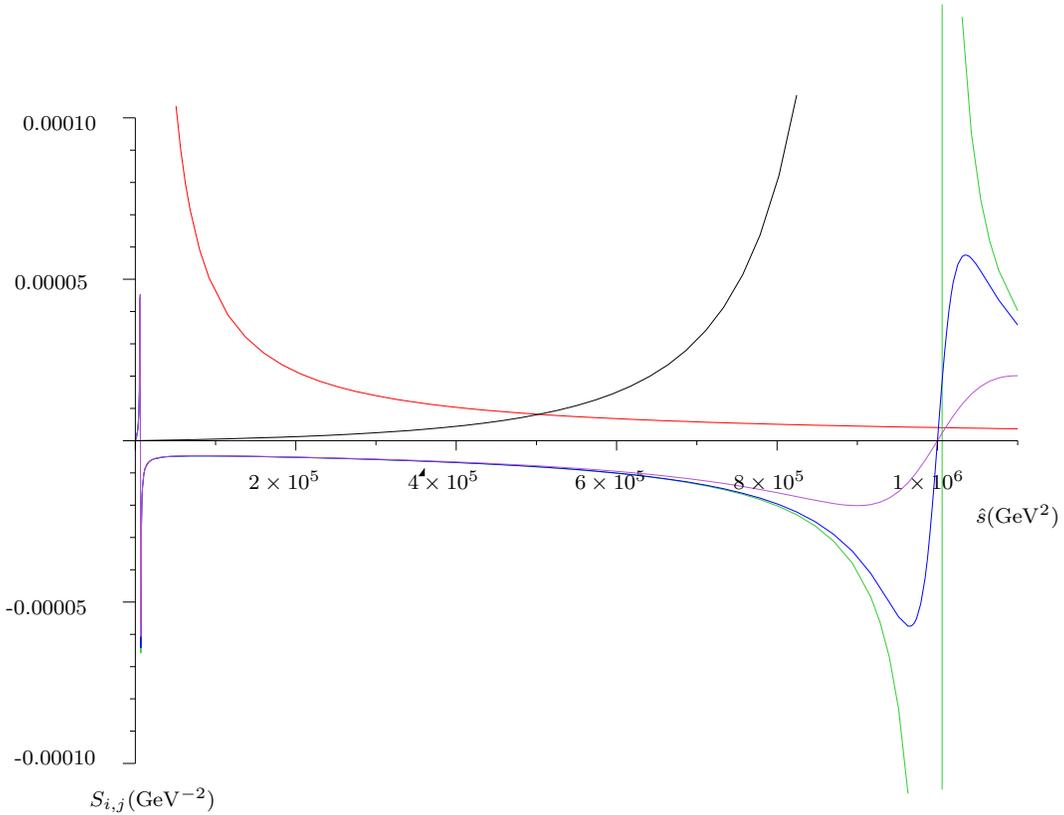}
  \end{center}
\caption{The interference term $S_{int}$ in the case where $h_{W'} = 1$, $M_{W'} = 1 \tev$, plotted against $\hat{s}$, for different widths: $\Gamma_{W'} = 1, 35, 100 \gev$ (green, blue, purple respectively). The terms $S_{W,W}$ (red) and $S_{W',W'}$ (black) are shown for comparison. It is evident that $S_{int}$ is negative in the intermediate region $(M_W^2,M_{W'}^2)$. It is also clear that the magnitude of the interference term is comparable to $S_{W,W}$ and $S_{W',W'}$. As the width decreases the negative peak becomes narrower, but there always exists a portion of the curve which is independent of the width.}
\label{fig:sint}
\end{figure}

We expect to observe interference effects in the differential cross-section simply because the Drell-Yan process $pp \rightarrow W/W' \rightarrow \ell \nu X$ can proceed either via a SM $W$ or a $W'$ in the reference model. The matrix element squared for the process may be decomposed in the following way:
\begin{equation}\label{eq:mesquared}
|\mathcal{M}|^2= |\mathcal{M}_{W}|^2 + |\mathcal{M}_{W'}|^2 + 2 \mathrm{Re}(\mathcal{M}_W^* \mathcal{M}_{W'})
\end{equation}
It is easy to see that the interference term depends on the functions $S(\hat{s})$ and $A(\hat{s})$ (eq.~\ref{eq:Sterm} and eq.~\ref{eq:Aterm}). Here we discuss the function $S(\hat{s})$ when studying interference effects, although the arguments for $A(\hat{s})$ are equivalent. Just as with the squared matrix element, $S(\hat{s})$ can be decomposed into pieces which are due to the $W$ and $W'$ individually and an interference piece:
\begin{equation}\label{eq:stermdecomp}
S = S_{W,W} + S_{W',W'} + S_{W,W'} + S_{W',W} = S_{W,W} + S_{W',W'} + 2 S_{int}
\end{equation} 
where we have defined the interference term $S_{int} \equiv S_{W,W'} = S_{W',W}$. Explicitly, this interference term may be written as:
\begin{equation}\label{eq:sint}
S_{int} = \left[\hat{s} \frac{ (\hat{s} - M_W^2) (\hat{s} - M_{W'}^2) + \Gamma_{W} \Gamma_{W'} M_{W} M_{W'}) } { [(\hat{s} - M_W^2)^2 + \Gamma_W^2 M_W^2][W\rightarrow W']}\right] (1 + h_W h_{W'})^2
\end{equation}
where we have set all the couplings $C^{\ell,q}_{W/W'} = 1$ and $k_W=k_{W'}=1$. It is evident that when $h_W = 1$ (SM) and $h_{W'} = -1$ then $S_{int} = 0$. Hence there is \textit{no} interference for the case of the SM $W$ and right-handed $W'$, and the square of the total matrix element for the process can be written as the sum of the squares of the individual matrix elements for the $W$ and $W'$:
\begin{equation}\label{eq:merhwp}
|\mathcal{M}(h_{W'} = -1)|^2= |\mathcal{M}_{W}|^2 + |\mathcal{M}_{W'}|^2
\end{equation}
This is what we would expect just by noting that the $W'$ decays to different final state particles than the $W$. 
However, when $h_W = 1$ and $h_{W'} = 1$, i.e. both left-handed, we have $S_{int} \ne 0$. In fact, by examination of the expression for $S_{i,j}$ (eq.~\ref{eq:Sterm}), we can observe that $S_{int}$ should be of the same order of magnitude as $S_{W',W'}$ and $S_{W,W}$. figure~\ref{fig:sint} shows the variation of the interference term for the case $M_{W'} = 1 \tev$ as well as $S_{W,W}$ and $S_{W',W'}$. We can observe that $S_{int}$ is negative in the intermediate mass squared region $\hat{s} \in (M_W^2,M_{W'}^2)$. The total cross-section in this region is reduced in comparison to the sum of the individual $W$ and $W'$ cross-sections. It is important to note that the interference term is non-vanishing and comparable in magnitude to the other terms in $S(\hat{s})$ even as $\Gamma_{W'}\rightarrow 0$, a clear indication of why NWA is not justified in the intermediate region. If $\Gamma_{W'} = 0$ the particles do not overlap directly with each other, however off-mass-shell effects still cause interference. We emphasize the fact that the interference is negative when a SM $W$ interferes with a left-handed $W'$. This leads to a reduction in the cross section with respect to the Standard Model expectation, a possibility seldomly considered in experimental searches.
\section{Extension to NLO}\label{sec:nlo}
Next, we extended the simulation to NLO using the {\tt MC@NLO} and the {\tt POWHEG}
methods. The {\tt MC@NLO} method has previously been applied to the
hadroproduction of gauge boson pairs \cite{Frixione:2002ik, Frixione:2002bd}, heavy quark-antiquark pairs
\cite{Frixione:2003ei} and single-top production \cite{Frixione:2005vw}. In these applications, the Fortran Monte Carlo event generator {\tt HERWIG} \cite{Corcella:2000bw} was used to simulate
the parton shower and hadronization. Within the framework of {\tt Herwig++}, it has been applied to $e^+ e^-$ annihilation to hadrons and Drell-Yan vector boson production,
\cite{LatundeDada:2007jg}.
 
The method is based upon a careful expansion of the NLO results,
in order to match certain features of the event generator used, in this case {\tt Herwig++}. The shower approximation to the NLO matrix element in {\tt Herwig++} must be
subtracted from the exact NLO result in order to avoid double counting. This subtraction
generates a number of negative weighted events which however, are few enough 
so that the number of events required for a smooth distribution is comparable to leading
order simulations. 


The {\tt POWHEG} method on the other hand generates the hardest emission of the parton
shower to NLO accuracy first and, for angular ordered showers such as {\tt Herwig++}, adds
a truncated shower of soft and wide angled emissions between the hard scale and the scale of the hardest emission. The
resulting partons are then showered subject to a $p_{\rm T}$ veto so that no subsequent emissions
have $p_{\rm T}$ greater than the hardest emission. This method has been applied to $Z$ pair
production \cite{Nason:2006hfa}, heavy flavour production \cite{Frixione:2007nw},
Drell-Yan vector boson production \cite{Alioli:2008gx,Hamilton:2008pd}, $e^+e^-$
annihilation into hadrons \cite{LatundeDada:2006gx} and into top pairs and their decays \cite{LatundeDada:2008bv}, and NLO Higgs boson production via gluon fusion \cite{Alioli:2008tz}.

The advantage of this method over the
{\tt MC@NLO} method is the independence of the procedure on the event generator used and
the generation of only positive weighted events. 


We now briefly discuss both methods and their application to $W'$ boson production. Full
details of the application of the {\tt MC@NLO} method to vector boson production can be found in Section $6$ of
\cite{LatundeDada:2007jg}. Details of the application of the {\tt POWHEG} method can be
found in Chapter $4$ of \cite{seyi} where vector boson production is discussed in
detail. The implementation of a truncated shower of at most one emission in the framework of {\tt Herwig++} is also discussed.

\subsection{The {\tt MC@NLO} method}
The NLO cross-section for the production of $W'$ bosons
can be written as a sum of two contributions,
\begin{equation}
\label{sigg}
\sigma_{\rm NLO} = \sigma_{q\bar{q}'} + \sigma_{(q,\bar{q}')g}
\end{equation}
where $\sigma_{q\bar{q}'}$ is the contribution from $q\bar{q}'$ annihilation and
$\sigma_{(q,\bar{q}')g}$ is the contribution from the Compton subprocesses. In the modified minimal subtraction (MSbar) factorization
scheme, these are:
\begin{eqnarray}
\label{qbg}
\sigma_{q\bar{q}'} &=& \sigma_0 \sum_q \int dx_1dx_2 \frac{x[D_q(x_1)D_{\bar{q}'}(x_2)+q\leftrightarrow \bar{q'}]}{D_q(x_q)D_{\bar{q}'}(x_{\bar{q}'})} \left[\delta(1-x)
    +\frac{\alpha_S}{2\pi} C_F \left \{
      -2\frac{1+x^2}{1-x}\ln x \right. \right. \nonumber \\
&+& \left. \left. 4(1+x^2)\left(\frac{\ln(1-x)}{1-x}\right)_++\left(-8+\frac{2\
}{3}\pi^2\right)\delta(1-x)\right\}\right] \nonumber \\
\sigma_{(q,\bar{q}')g} &=& \sigma_0 \sum_{q,\bar{q}'} \int dx_1dx_2\frac{x[D_{(q,\bar{q}')}(x_1)D_g(x_2)+ (q,\bar{q}')\leftrightarrow
g]}{D_q(x_q)D_{\bar{q}'}(x_{\bar{q}'})}\frac{\alpha_S}{2\pi}T_R\left [
\frac{1}{2}+3x-\frac{7}{2}x^2 \right.
 \nonumber \\
&+& \left.(x^2+(1+x^2))\ln\frac{(1-x)^2}{x}\right] \,,
\end{eqnarray}
where $\sigma_0$ is the Born differential cross-section $\frac{d^2\sigma_0}{dQ^2dY}$ with $Q$ the invariant mass and $Y$ the rapidity of the vector boson. The $x_1,x_2$ are the NLO momentum fractions and
$x_q,x_{\bar{q}'}$ are the Born momentum fractions with $Q^2 = x_qx_{\bar{q}'}S$, if $S$ denotes the hadronic centre-of-mass energy. Also, $x = \frac{x_qx_{\bar{q}'}}{x_1x_2}$ and $D_q(x_1) = x_1f_q(x_1)$ etc., with $f_q(x_1)$ being the parton distribution
function of parton $q$. 

Focusing on the $q\bar{q}'$ annihilation process for the moment, if we introduce the variable
\begin{equation}
y = \cos \theta,
\end{equation}
where $\theta$ is the scattering angle of the emitted parton in the partonic COM frame, we can
re-write $\frac{\sigma_{q\bar{q}'}}{\sigma_0}$ as an integral over $x$ and $y$:
\begin{eqnarray}
\label{sub}
\frac{\sigma_{q\bar{q}'}}{\sigma_0} &=& \sum_q \int dxdy \left[\left\{
\frac{x[D_q(x_1)D_{\bar{q}'}(x_2)+q\leftrightarrow \bar{q'}]}{D_q(x_q)D_{\bar{q}'}(x_{\bar{q}'})}\frac{1}{2} \left(\delta(1-x)
    +\frac{\alpha_S}{2\pi} C_F \left (-2
      \frac{1+x^2}{1-x}\ln x \right. \right.\right. \right. \nonumber \\
&+&
\left. \left. \left.
  \left. 4(1+x^2)\left(\frac{\ln(1-x)}{1-x}\right)_++\left(-8+\frac{2}{3}\pi^2\right)\delta(1-x)\right)\right) -M_{q\bar{q}'}(x,y) \right\}+ M_{q\bar{q}'}(x,y) \right] \nonumber \\
\end{eqnarray}
where $M_{q\bar{q}'}(x,y)$ is the real emission matrix element. Since we have subtracted this
contribution from the total cross-section, in the curly brackets we are left with the sum of the Born, virtual
and QCD PDF correction contributions. Now we can define an infrared-safe observable $O$
whose NLO expectation value is given by:
\begin {eqnarray}
\langle O^{q\bar{q}'} \rangle&=&\sum_{q}\int
dxdy\left[O_{W'}\left\{\frac{x[D_q(x_1)D_{\bar{q}'}(x_2)+q\leftrightarrow
\bar{q}]}{D_q(x_q)D_{\bar{q}'}(x_{\bar{q}'})}\frac{1}{2}\left (\delta(1-x)
  \right. \right.\right.\nonumber \\
&+&\left.\left.\frac{\alpha_S}{2\pi}C_F\left(-2\frac{1+x^2}{1-x}\ln x+4(1+x^2)\left(\frac{\ln(1-x)}{1-x}\right)_+
      \right.\right.\right. \nonumber \\
&+& \left. \left. \left.\left.\left(-8+\frac{2}{3}\pi^2\right)\delta(1-x)\right)\
\right)-M_{q\bar{q}'}(x,y)\right\}+O_{W'g}M_{q\bar{q}'}(x,y)\right] \,,
\end{eqnarray}
where $O_{W'}$ and $O_{W'g}$ are observables arising from hadronic final states generated from  $q+\bar{q} \rightarrow W'$ and
$q+\bar{q} \rightarrow W'+g$ starting configurations respectively. This however is not
entirely correct because of double counting in the final states represented by $O_{W'}$
which are already included in the states arising from $O_{W'g}$. The solution to this is
the subtraction of the parton shower contributions which we denote $M_{C_{q\bar{q}'}}(x,y)$
from the regions in which the parton shower contributes (the jet region $J$) and integrate the full matrix
element in the hard emission region $D$, left untouched by the shower. This gives for
$\langle O^{q\bar{q}'} \rangle$:
\begin {eqnarray}
\langle O^{q\bar{q}'} \rangle&=&\sum_{q}\int_J\left[O_{W'}\left\{\frac{x[D_q(x_1)D_{\bar{q}'}(x_2)+q\leftrightarrow
\bar{q}']}{D_q(x_q)D_{\bar{q}'}(x_{\bar{q}'})}\frac{1}{2}\left (\delta(1-x)+\frac{\alpha_S}{2\pi}C_F\left(-2\frac{1+x^2}{1-x}\ln x
  \right. \right. \right.\right.\nonumber \\
&+&\left.\left.\left.4(1+x^2)\left(\frac{\ln(1-x)}{1-x}\right)_+
+\left(-8+\frac{2}{3}\pi^2\right)\delta(1-x)\right)\right)-M_{q\bar{q}'}+M_{{C}_{q\bar{q}'}}\right\}\nonumber \\
&+&\left.O_{W'g}\left \{M_{q\bar{q}'}-M_{{C}_{q\bar{q}'}}\right\}\right] \nonumber\\
&+&\sum_{q}\int_D \left[O_{W'}\left\{\frac{x[D_q(x_1)D_{\bar{q}'}(x_2)+\
q\leftrightarrow
\bar{q}']}{D_q(x_q)D_{\bar{q}'}(x_{\bar{q}'})}\frac{1}{2}\left (\delta(1-x)+\frac{\
\alpha_S}{2\pi}C_F\left(-2\frac{1+x^2}{1-x}\ln x
  \right. \right. \right.\right.\nonumber \\
&+&\left.\left.\left.4(1+x^2)\left(\frac{\ln(1-x)}{1-x}\right)_++\left(-8+\frac{2}{3}\pi^2\right)\delta(1-x)\right)\right)-M_{q\bar{q}'}\right\}\nonumber \\
&+&\left.O_{W'g}M_{q\bar{q}'}\right] \;.
\end{eqnarray}
A similar functional $\langle O^{(q,\bar{q}')g} \rangle$ can be generated for the Compton
subprocesses. Events can then be generated in the different regions of phase space according
to their contributions to the above integrals. These events are then interfaced with
\texttt{Herwig++} and showered. Full details of the algorithm for event generation can be found in
\cite{LatundeDada:2007jg}.
\subsection{The {\tt POWHEG} method}
This method as described in \cite{powheg} involves the generation of the hardest radiation from
the parton shower according to the real emission matrix element and independently of the
shower Monte Carlo generator used. If we introduce:
\begin{equation}
\centering
\label{eq:M}
R_{v,r}=M_{q\bar{q}'}+M_{(q,\bar{q}')g} \,,
\end{equation}
where $M_{q\bar{q}}$ and $M_{(q,\bar{q}')g}$ are real emission matrix elements for $q\bar{q}'$
annihilation and the Compton subprocesses respectively, we can write the cross section for the hardest gluon emission event as:
\begin{equation}
\centering
\label{eq:sig2}
d \sigma=\sum_q \bar {B}^q_{v} d \Phi_{v}\left[\Delta^q(0)+\Delta^q(p_{\rm T})R_{v,r}d
  \Phi_{r}\right] \;.
\end{equation}
The index $q$ runs over all quarks and anti-quarks. The subscript $v$ represents the Born
variables, which in this case are the invariant mass $Q$ and
the rapidity $Y$ of the boson, $r$ represents the
radiation variables $x, y$ and $d \Phi_{v}, d\Phi_{r}$ are the Born and real emission phase spaces respectively.

$\Delta^q(p_{\rm T})$ is the modified Sudakov form factor for the hardest emission with
transverse momentum $p_{\rm T}$, as indicated by the Heaviside function in the exponent of
eq.~\ref{eq:dnloo}:
\begin{equation}
\centering
\label{eq:dnloo}
\Delta^q(p_{\rm T})=\exp \left[-\int d\Phi_{r} R_{v,r}\Theta(k_{\rm T}(v,r)-p_{\rm T})\right]\;.
\end{equation}
where $k_{\rm T}$ is the transverse momentum of
the hardest emission relative to the splitting axis and in this case is given by:
\begin{equation}
\centering
\label{eq:kT2}
k_{\rm T}(x,y)=\sqrt{\frac{Q^2}{4x}(1-x)^2(1-y^2)} \;.
\end{equation}
Furthermore,
\begin{equation}
\centering
\label{eq:Bbarr}
\bar{B}^q_{v}=B^q_{v}+V^q_{v}+\int (R_{v,r}-C_{v,r})d \Phi_{r}\;.
\end{equation}
$\bar{B}^q_{v}$ is the sum of the Born, $B^q_{v}$, virtual, $V^q_{v}$ and real, $R_{v,r}$ terms, (with some
counter-terms, $C_{v,r}$). The Born variables are generated with distribution
$\bar{B}^q_{v}$, with the radiation variables of the first emission generated according to $[\Delta^q(0)+\Delta^q(p_{\rm T})R_{v,r}d \Phi_{r}]$.

In the MSbar factorization scheme, the contribution to the order $\alpha_S$
cross-section for $W'$ production is given in eq.~\ref{sigg} and eq.~\ref{qbg}.
The function $\bar{B}^q$ in eq.~\ref{eq:Bbarr} can then be written down as a
sum of finite terms using the subtraction method. In this paper, we borrow the MC@NLO
subtraction formula introduced in 
eq.~\ref{sub} and write a function $\tilde{B}^q(Q^2,Y)$ as:
\begin{eqnarray}
\label{Bbar2}
\tilde{B}^q(Q^2,Y)&=&\sum_{q} \int dxdydQ^2dY \, \frac{d^2\sigma_0}{dQ^2dY}\left[ \
\frac{x[D_q(x_1)D_{\bar{q}'}(x_2)+q\leftrightarrow
\bar{q}]}{D_q(x_q)D_{\bar{q}'}(x_{\bar{q}})}\frac{1}{2}\left[\delta(1-x)
  \right. \right.\nonumber \\
&+&\frac{\alpha_S}{2\pi}C_F\left\{-2\frac{1+x^2}{1-x}\ln x\left.+4(1+x^2)\left(\frac{\ln(1-x)}{1-x}\right)_+
  \right. \right.
\nonumber \\
&+& \left. \left.\left(-8+\frac{2}{3}\pi^2\right)\delta(1-x)\right\}\right]-M_{q
  \bar{q}} +M_{C_{q \bar{q}'}}+\left\{M_{q\bar{q}'} -M_{C_{q \bar{q}'}}\right\}\nonumber \\
&+& \frac{x[D_{(q, \bar{q}')}(x_1)D_g(x_2)+ (q,\bar{q}')  \leftrightarrow
g]}{D_q(x_q)D_{\bar{q}}(x_{\bar{q}})}\frac{\alpha_S}{2\pi}T_F\frac{1}{2}\left [\
\frac{1}{2}+3x - \frac{7}{2}x^2 \right.
 \nonumber \\
&+&\left.\left. (x^2+(1+x^2))\ln\frac{(1-x)^2}{x}\right] 
-M_{(q,\bar{q}')g} +M_{C_{(q,\bar{q}')g}} +\left\{M_{(q,\bar{q}')g}-M_{C_{(q,\bar{q}') g}}\right\}\right] \,, \nonumber \\
\end{eqnarray}
where we have written the virtual and PDF corrections in terms
of the real emission matrix elements and $M_C$ are the subtracted parton shower approximation terms in
the \texttt{Herwig++} jet regions. Note that the above prescription does not imply that the {\tt POWHEG} method depends on the shower MC used. We have simply used the shower approximation terms to defne a subtraction scheme for the definition of the NLO cross-section.

The flavour of the event, the Born variables $Q^2$ and $Y$, as well as radiation variables
$x$ and $y$ are then generated
according to the integrand in eq.~\ref{Bbar2}. The radiation variables are ignored which
amounts to integrating away these variables leaving the Born variables distributed
according to $\bar{B}^q(Q^2,Y)$. The radiation variables $x,y$ are generated according to:
\begin{equation}
\Delta^q(p_{\rm T})R(x,y)dxdy \;.
\end{equation}
Details of the algorithm used can be found in \cite{seyi}.
\section{Results}\label{sec:results}
\begin{figure}
  \vspace{1cm}
  \centering 
    \hspace{3.5cm}
    \includegraphics[scale=0.50, angle=90]{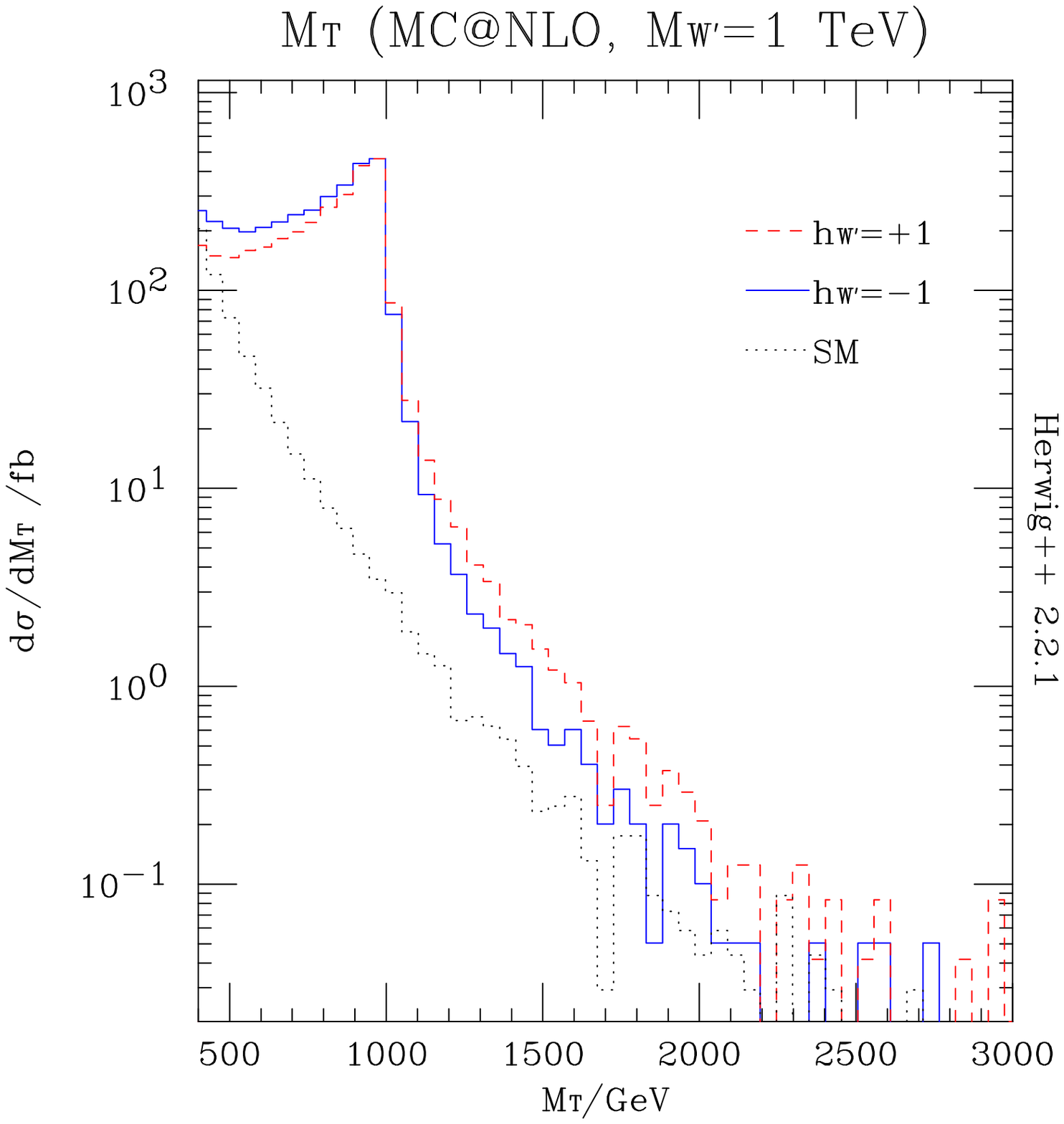}
    \hspace{5.0cm}
   \includegraphics[scale=0.50, angle=90]{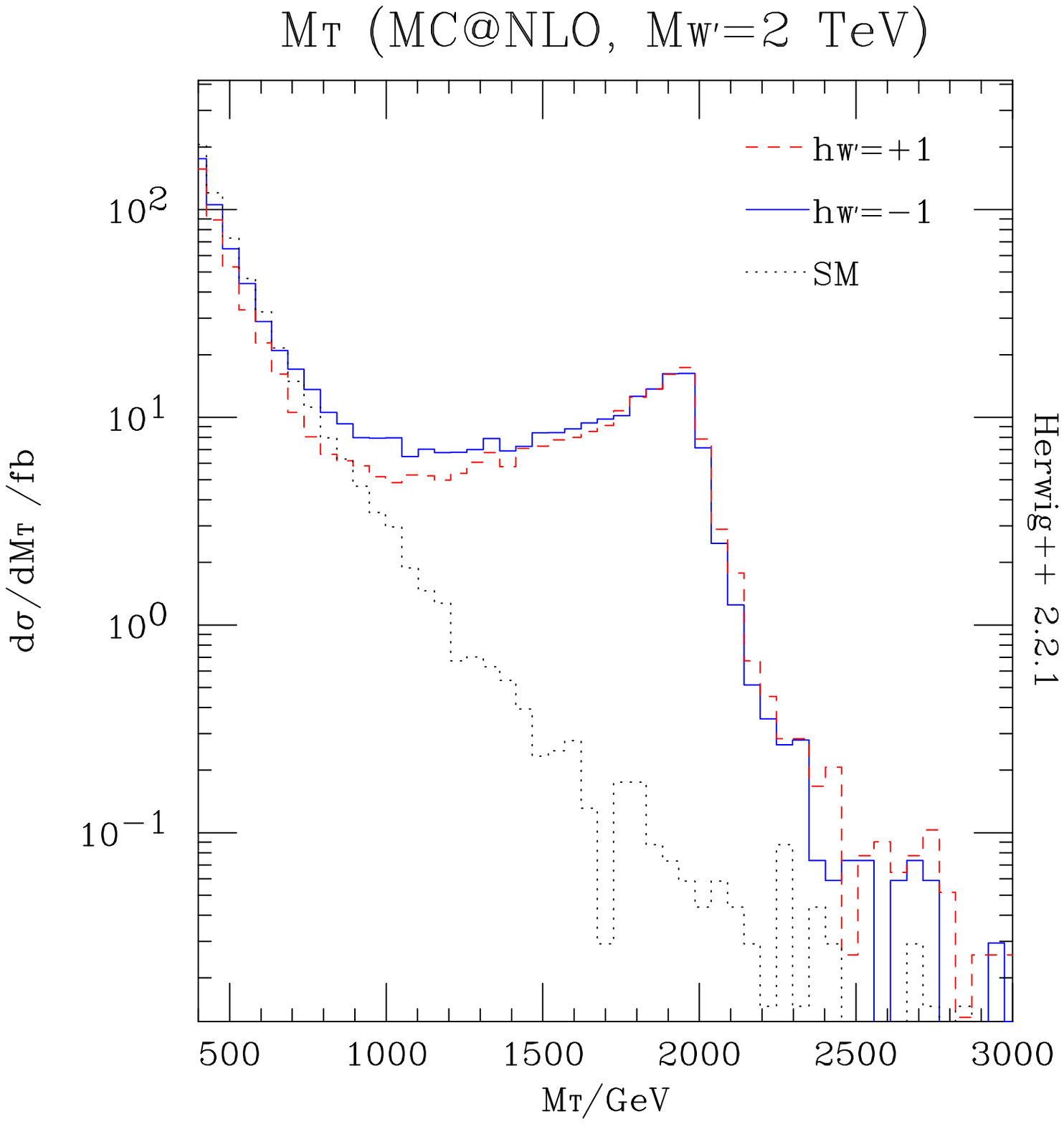}
   \vspace{0.75cm}
\caption{Transverse mass distributions at the LHC obtained using the \texttt{MC@NLO/Herwig++} method (cteq5m/MSbar), \texttt{POWHEG} (cteq5m) for purely left- and right-handed $W'$'s. The invariant mass range was taken to be $(0.4-3.0)\tev$. The plots correspond to masses/widths equal to $(1\tev,36\gev)$ and $(2\tev,72\gev)$. The significance of the destructive interference can be observed in the left-handed case; in the right-handed case the distribution is just the sum of the standard model $W$ and right-handed $W'$ contributions. Note that the plots are normalised to the NLO cross-section for each process.}
\label{fig:mt_mcnlo_LR}
\end{figure}

We present a sample of distributions of variables obtained for $\sim10^5$ events using the \texttt{Wpnlo} event generator, both at leading and next-to-leading order, using the \texttt{MC@NLO} (with \texttt{Herwig++}) and \texttt{POWHEG} methods. The general purpose event generator \texttt{Herwig++}, version 2.2.1~\cite{Bahr:2008tx}, was used. The $K$-factor (where $K=\sigma_{NLO} / \sigma_{LO}$) for the considered invariant mass range and for factorisation/renormalisation scales set to the default NLO scale $\mu_0 = \sqrt{k_T^2 + Q^2}$ (where $k_T$ and $Q$ are the dilepton transverse momentum and invariant mass respectively) was found to be $K\approx1.3$, in all cases. The plots have been normalised to unity (apart from figure~\ref{fig:mt_mcnlo_LR}) to emphasise the differences in the shape of the distributions.

\begin{figure}[!htb]
  \vspace{3.5cm}
  \centering 
    \includegraphics[scale=0.55, angle=90]{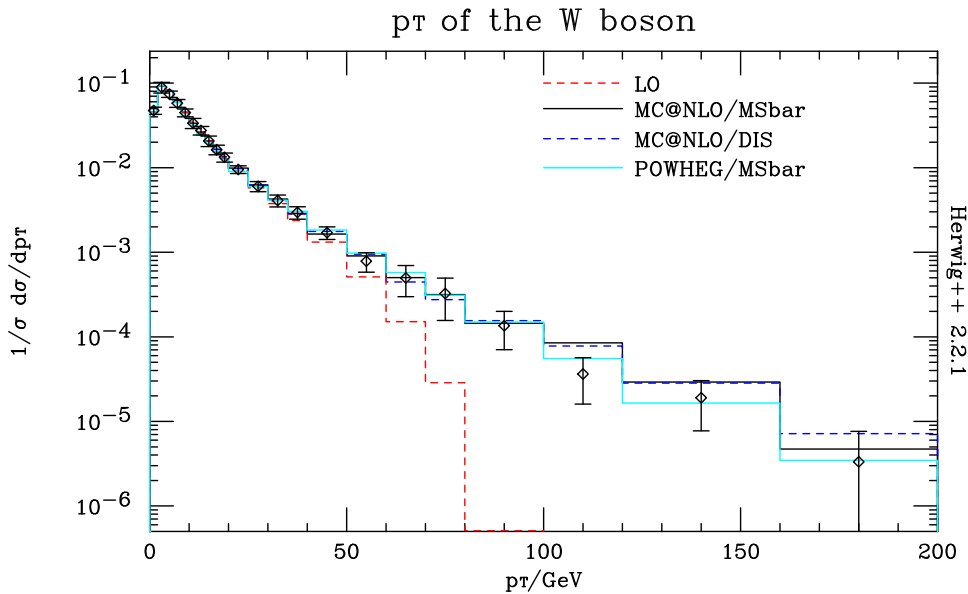}
\caption{Transverse momentum distribution at the Tevatron obtained for \texttt{MC@NLO/Herwig++} in the DIS and MSbar factorisation schemes (PDFs: cteq5d and cteq5m~\cite{Lai:1999wy} respectively), \texttt{POWHEG} MSbar (cteq5m) and LO (PDF: MRST2001LO~\cite{Martin:2001es}), in the mass range $(70-90)\gev$.}
\label{fig:tvt_pt}
\end{figure}

For validation purposes, figure~\ref{fig:tvt_pt} presents a comparison of the $W$ boson transverse momentum distribution, (assuming no $W'$) between Tevatron data (taken from~\cite{Affolder:1999jh}) and the three possible methods: leading order, \texttt{MC@NLO/Herwig++} and \texttt{POWHEG}. The plots include events in the invariant mass range $(70-90)\gev$. The \texttt{MC@NLO/Herwig++} and \texttt{POWHEG} distributions are evidently in agreement with the data within the statistical Monte Carlo and experimental uncertanties. The leading order $p_T$ distribution is cut off at the $W$ mass since this provides the only relevant scale in the shower, whereas the \texttt{MC@NLO/Herwig++} and \texttt{POWHEG} distributions extend to higher transverse momentum. 

\begin{figure}
  \centering 
  \hspace{4cm}
    \includegraphics[scale=0.60, angle=90]{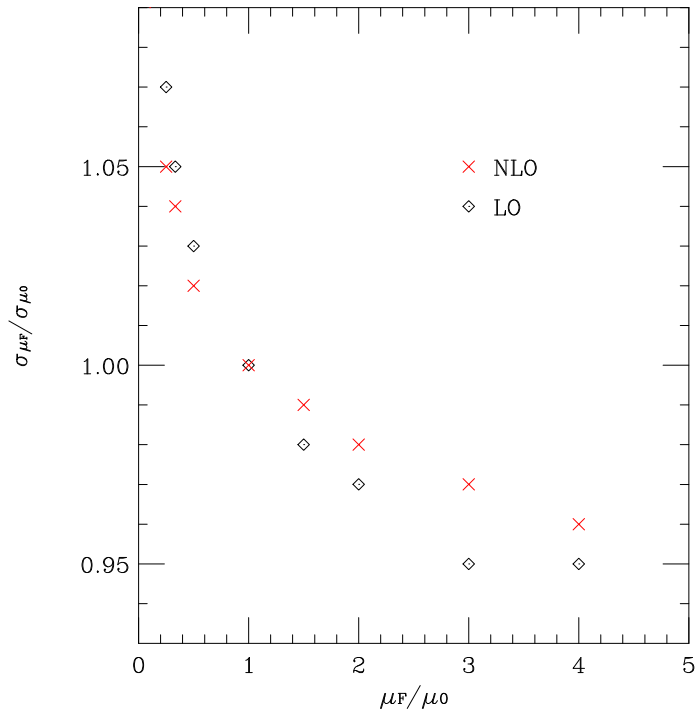}
    \vspace{1cm}
\caption{The normalised variation with scale of the cross section calculations at NLO (red crosses) and LO (black circles) are shown for a proton-proton collider at 14 TeV, $M_{W'} = 1\tev$, $\Gamma_{W'} = 36\gev$ and left-handed chirality in the invariant mass range [400,5000] GeV.}
\label{fig:scale}
\end{figure}

\begin{figure}[!htb]
  \centering 
  \vspace{1.0cm}
    \includegraphics[scale=0.33, angle=90]{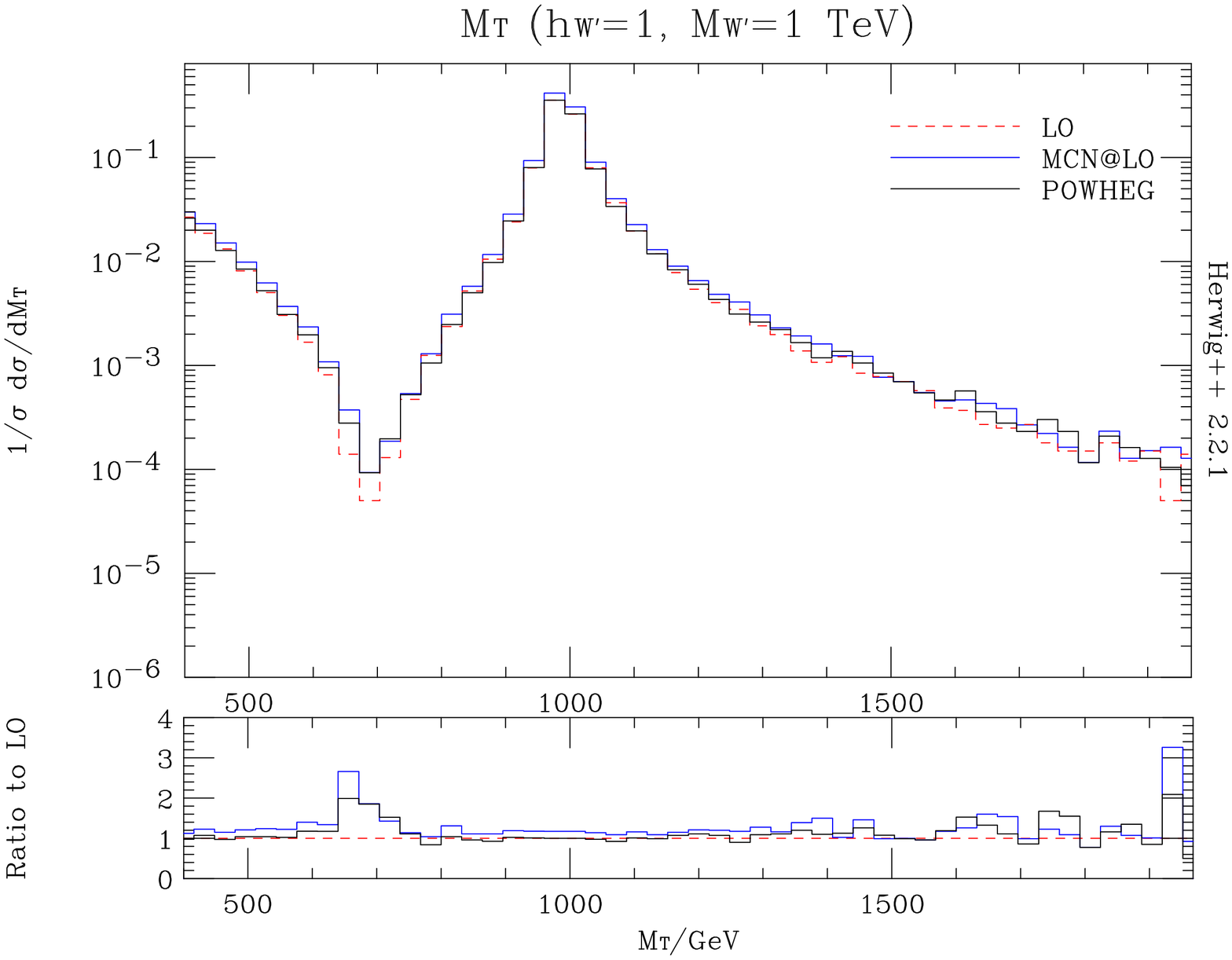}
    \hspace{1.0cm}
   \includegraphics[scale=0.33, angle=90]{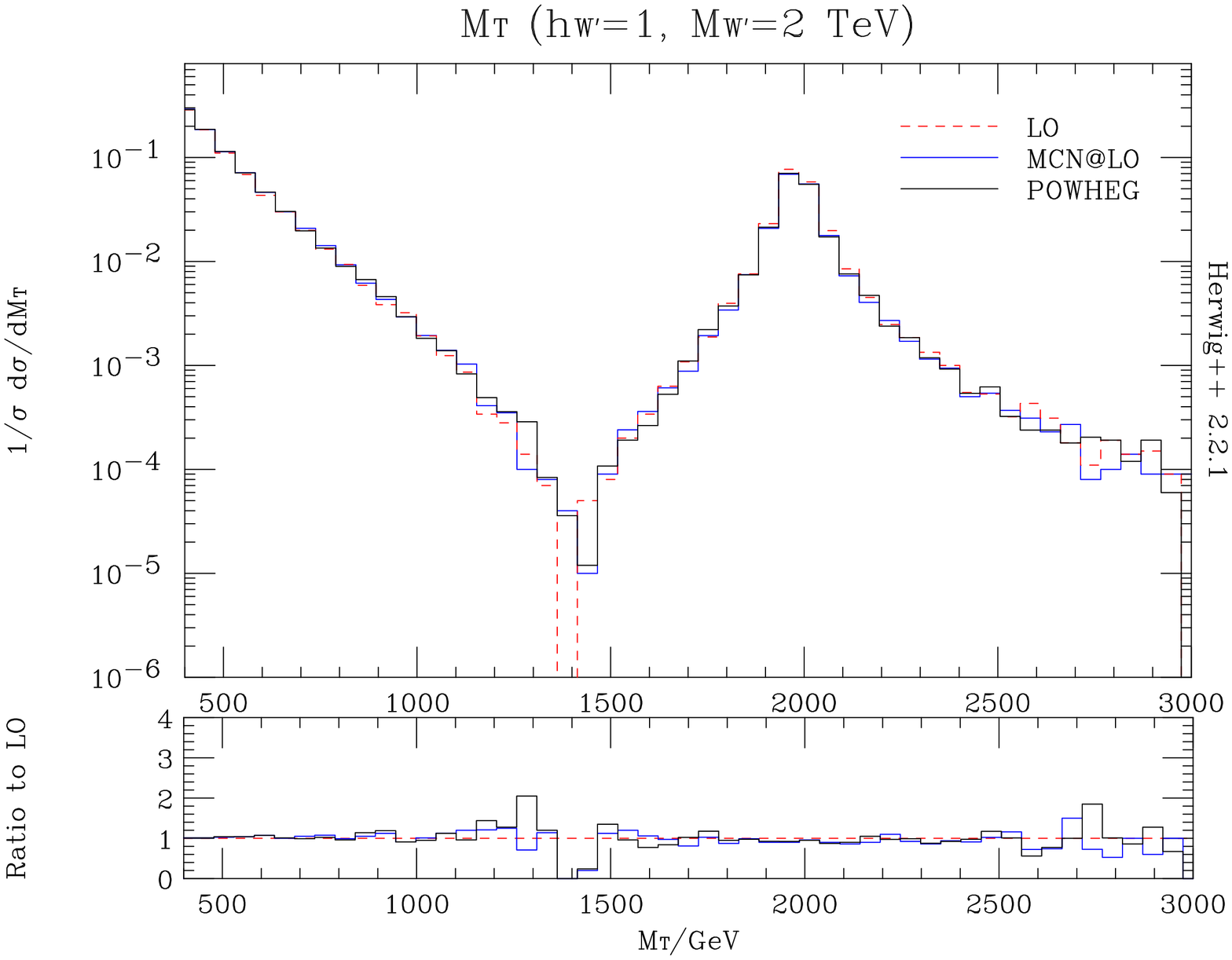}
\caption{Transverse mass distributions at the LHC obtained using the \texttt{MC@NLO/Herwig++} and \texttt{POWHEG} methods (cteq5m/MSbar) and LO (MRST2001LO) for a purely left-handed $W'$. The plots correspond to masses/widths equal to $[1\tev,36\gev]$ (left) and $[2\tev,72\gev]$ (right). The invariant mass range was taken to be $(0.4-3.0)\tev$ for the $1\tev$ case and $(0.4-5.0)\tev$ for the $2\tev$ case. The effect of the destructive interference can be observed in both cases. Note that the plots have been normalised to unity.}
\label{fig:mt_lh}
\end{figure}

\begin{figure}[!htb]
  \centering 
  \vspace{0.75cm}
  \includegraphics[scale=0.33, angle=90]{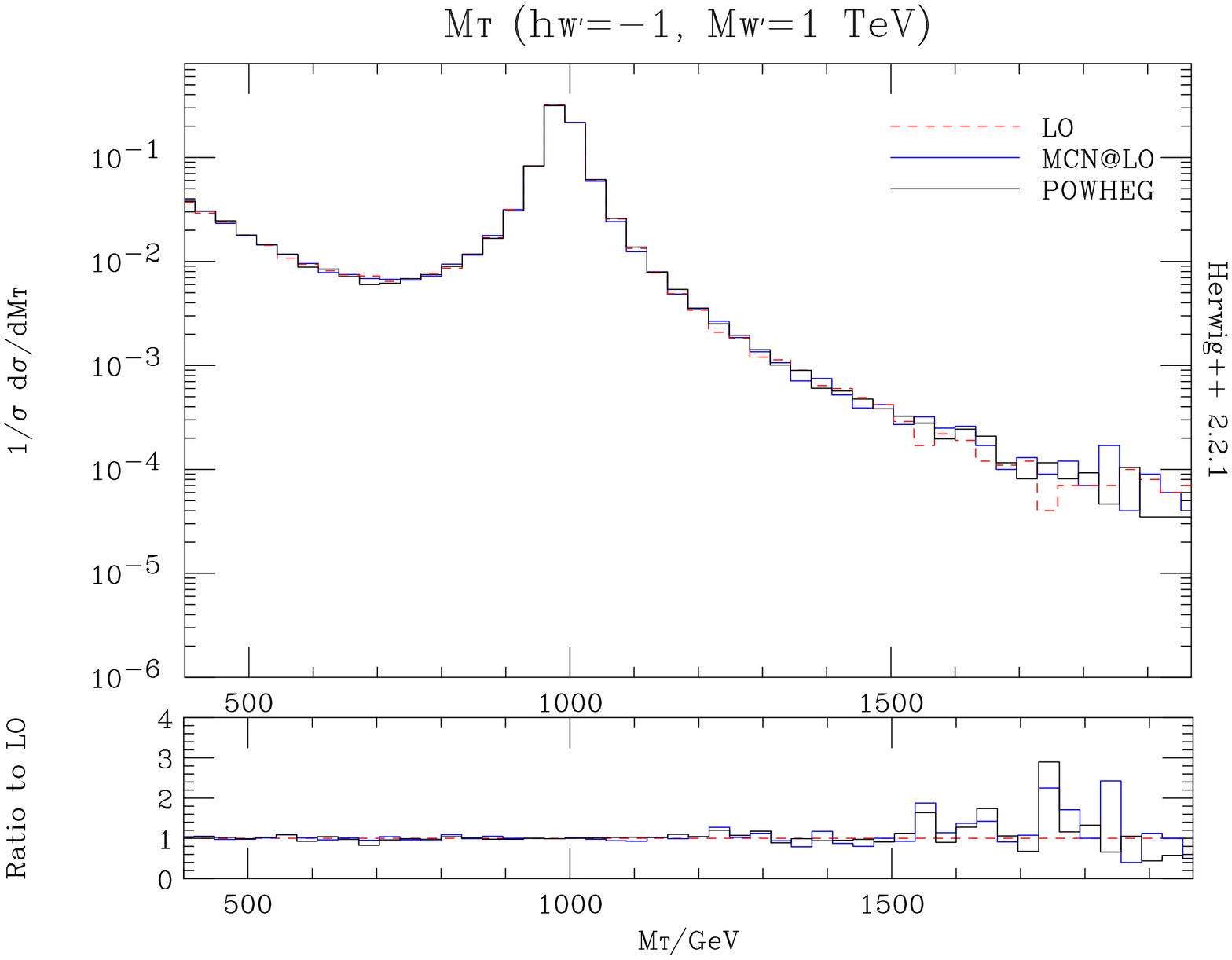}
    \hspace{1.0cm}
   \includegraphics[scale=0.33, angle=90]{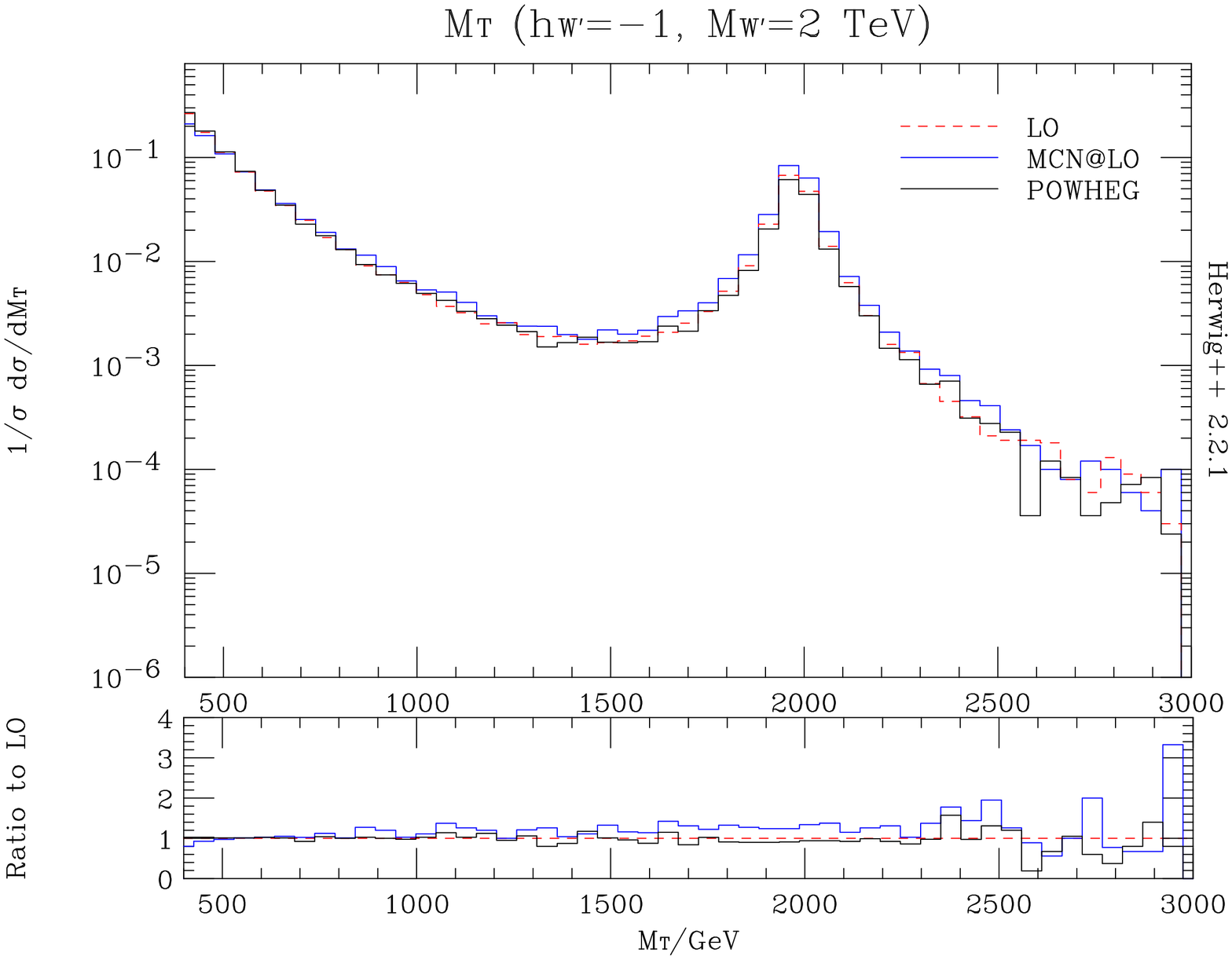}
    \caption{Transverse mass distributions at the LHC obtained using the \texttt{MC@NLO/Herwig++} and \texttt{POWHEG} methods (cteq5m/MSbar) and LO (MRST2001LO) for a purely right-handed $W'$. The invariant mass range and $W'$ mass and widths are identical to those in the previous figure.}
\label{fig:mt_rh}
\end{figure}

\begin{figure}[!htb]
  \vspace{0.75cm}
  \centering 
   \includegraphics[scale=0.33, angle=90]{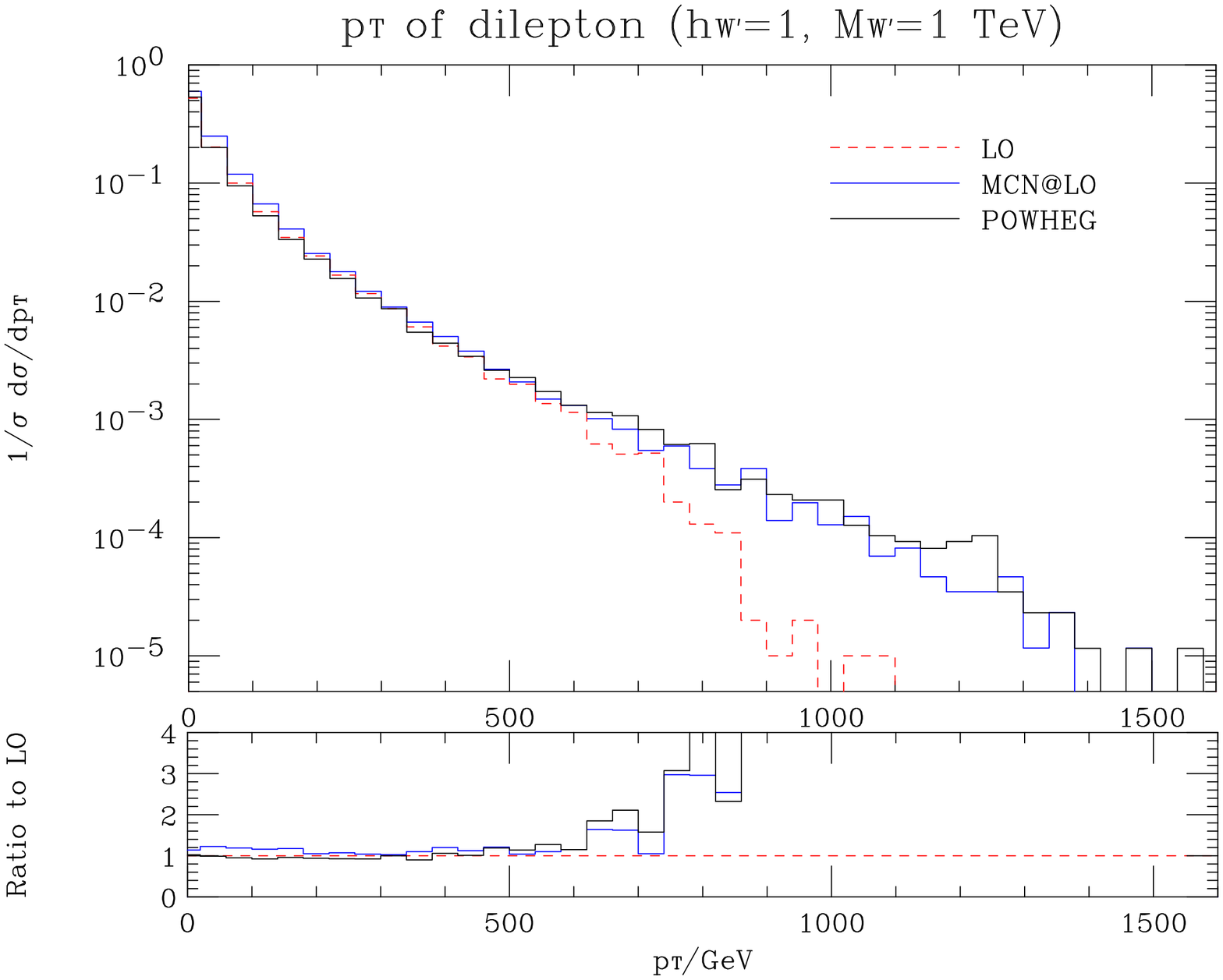}
    \hspace{1.0cm}
   \includegraphics[scale=0.33, angle=90]{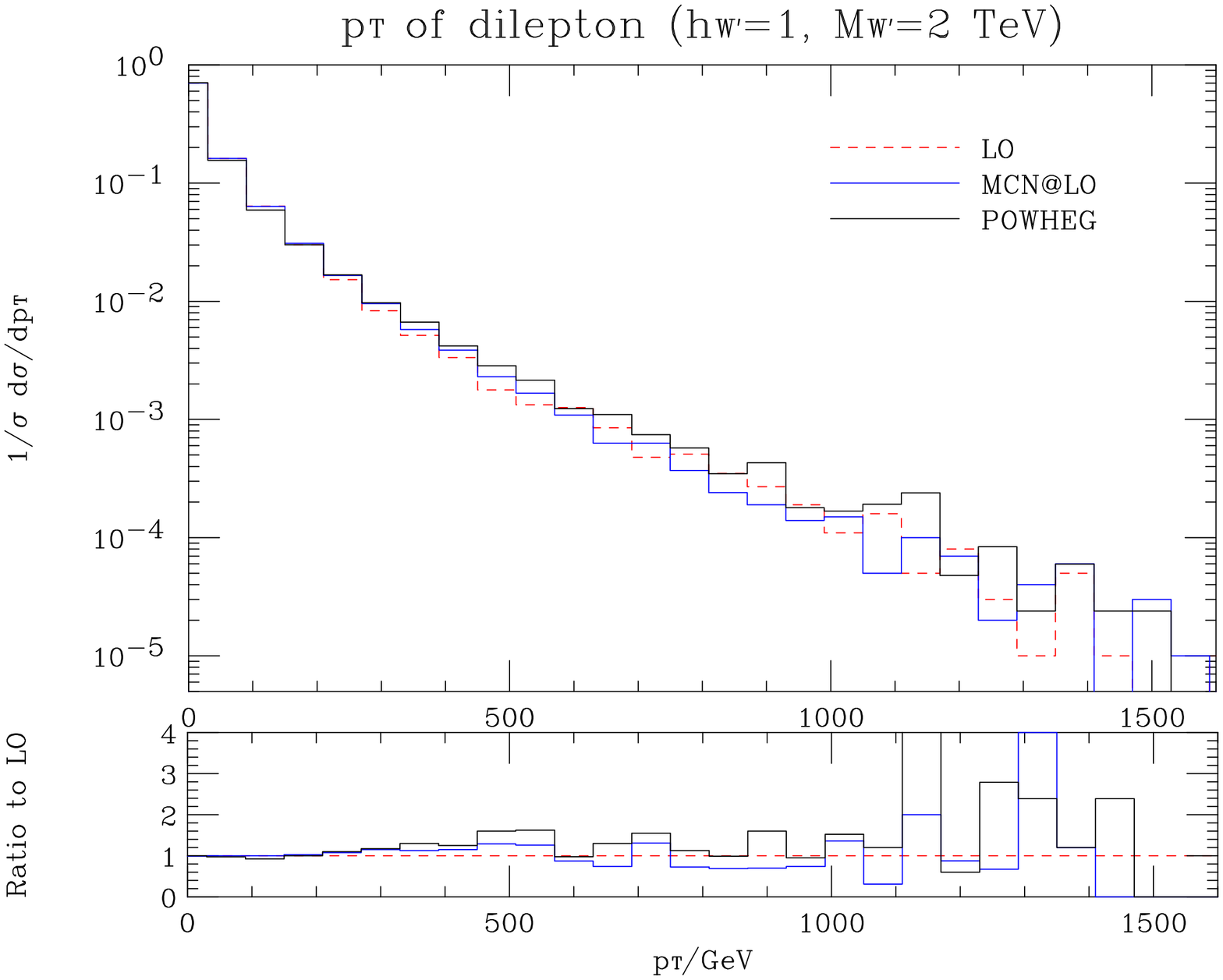}
\caption{Transverse momentum distributions at the LHC obtained using the \texttt{MC@NLO/Herwig++} and \texttt{POWHEG} methods (cteq5m/MSbar) and LO (MRST2001LO) for a purely left-handed $W'$. The invariant mass range and $W'$ mass and widths are identical to those in the previous figures.}
\label{fig:wpt_lh}
\end{figure}

\begin{figure}[!htb]
  \centering 
    \vspace{0.75cm}
    \includegraphics[scale=0.33, angle=90]{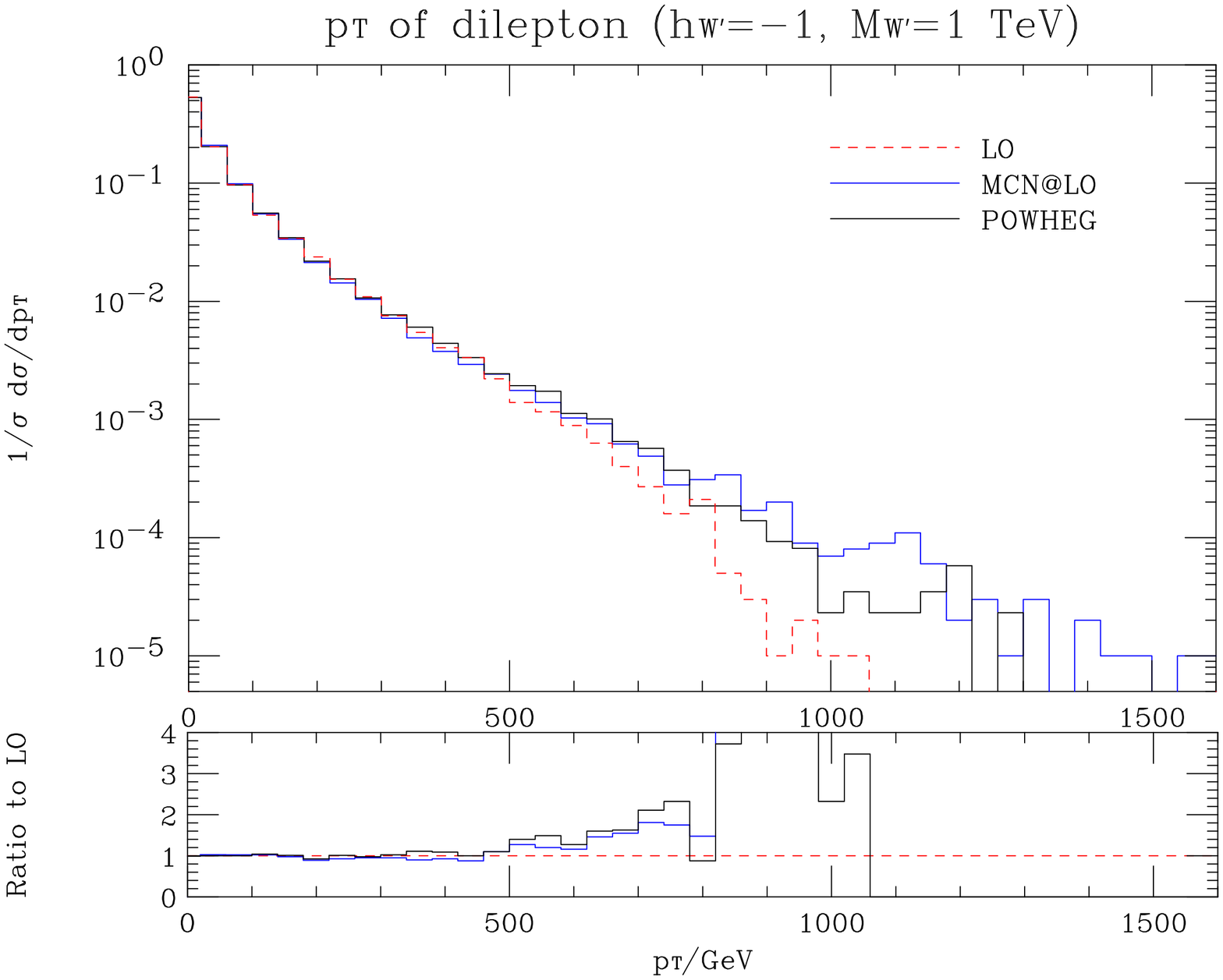}
    \hspace{1.0cm}
    \includegraphics[scale=0.33, angle=90]{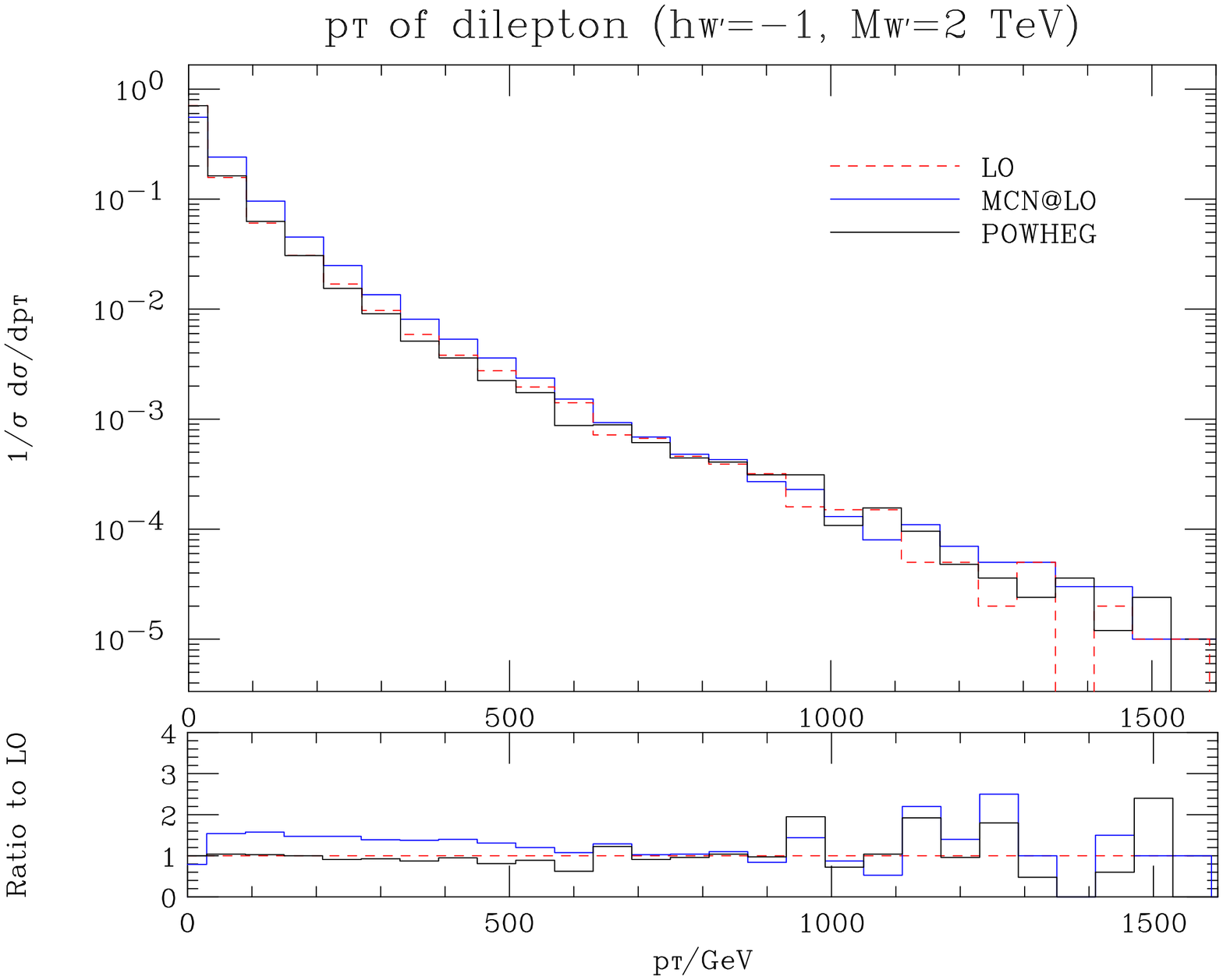}
\caption{Transverse momentum distributions at the LHC obtained using the \texttt{MC@NLO/Herwig++} and \texttt{POWHEG} methods (cteq5m/MSbar) and LO (MRST2001LO) for a purely right-handed $W'$. The invariant mass range and $W'$ mass and widths are identical to those in the previous figures.}
\label{fig:wpt_rh}
\end{figure}

\begin{figure}[!htb]
  \centering 
    \includegraphics[scale=0.33, angle=90]{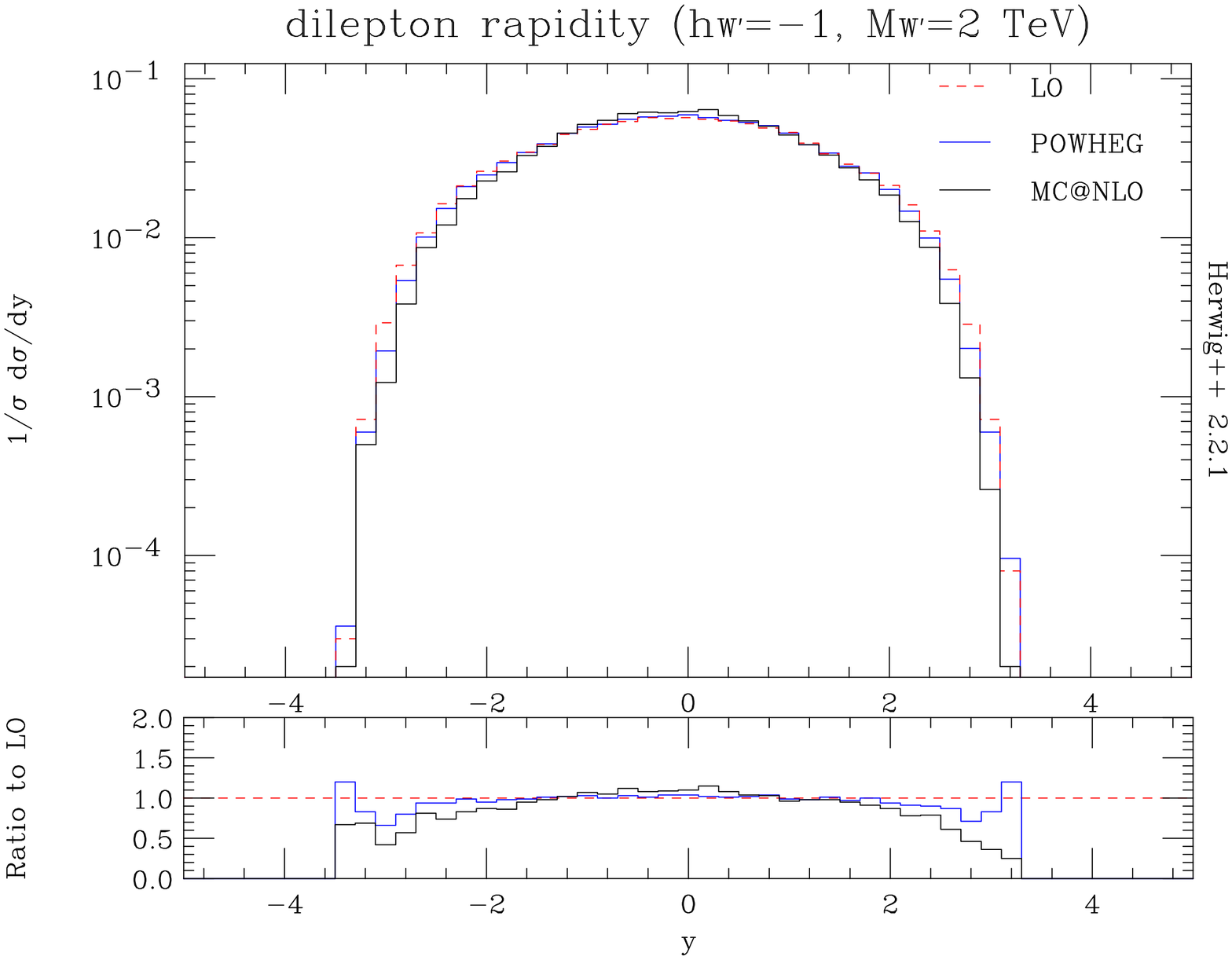}
    \hspace{1.0cm}
    \vspace{1.0cm}
    \includegraphics[scale=0.33, angle=90]{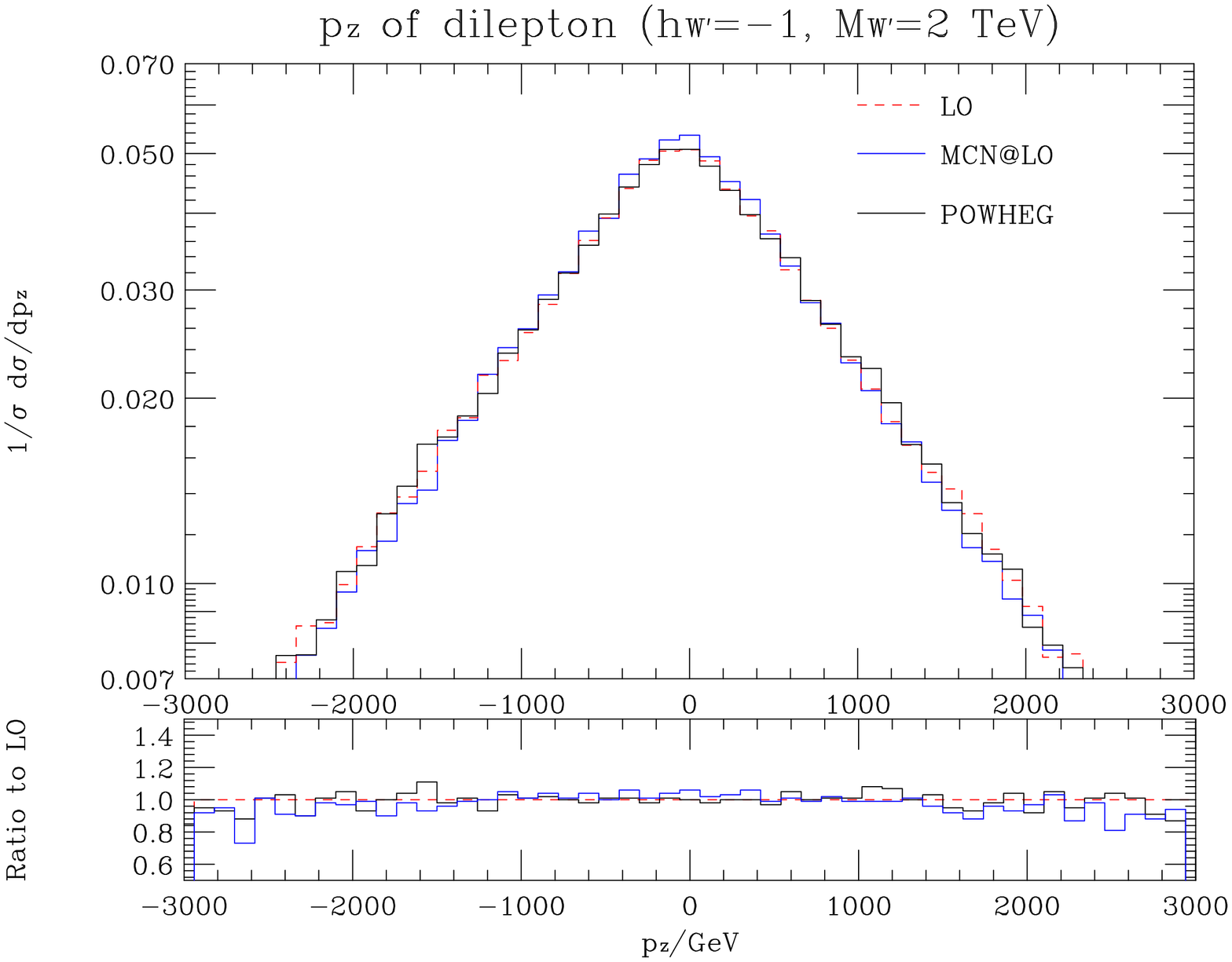}
    \hspace{4.0cm}    
    \includegraphics[scale=0.33, angle=90]{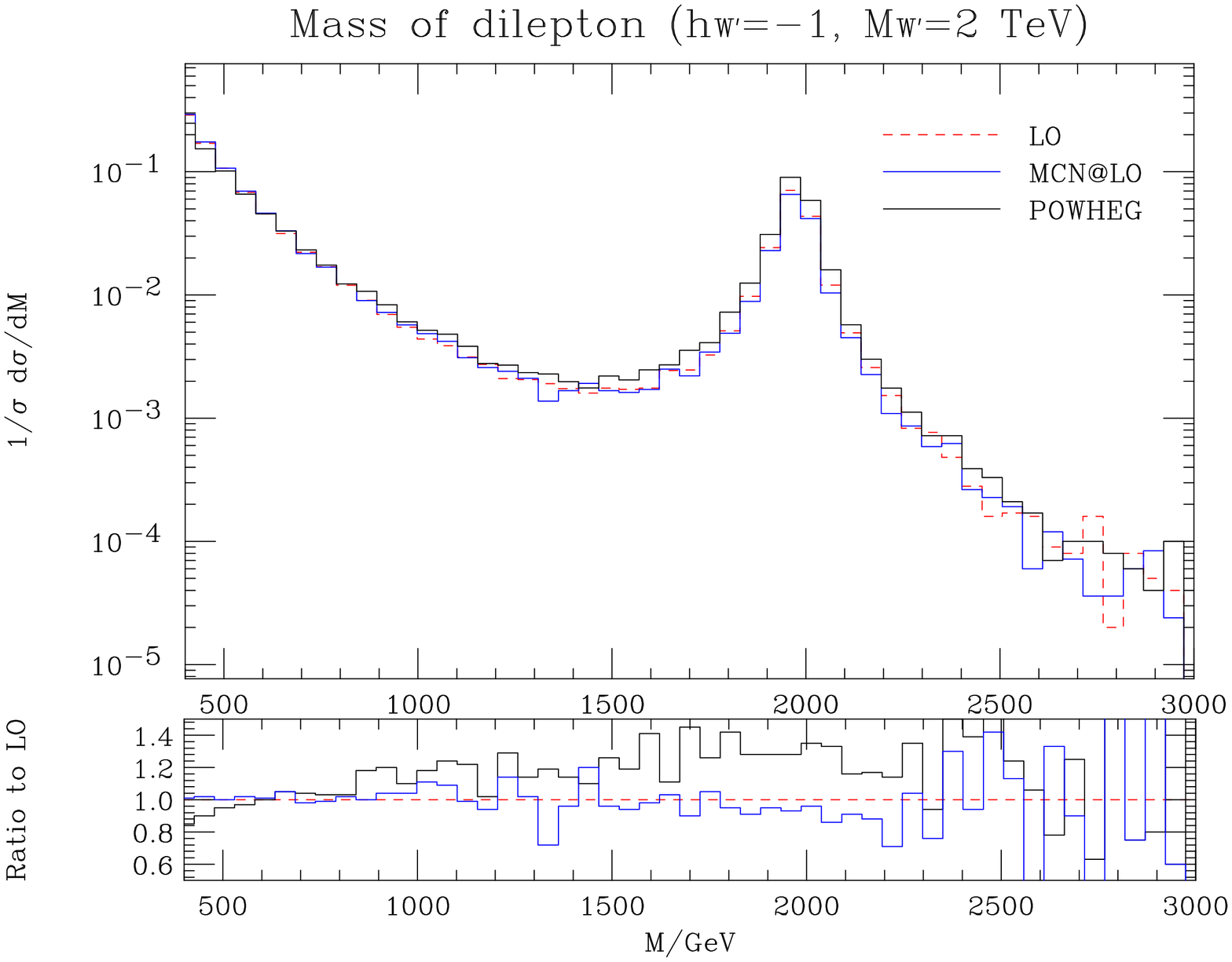}
    \caption{$W/W'$ rapidity (top left), longitudinal momentum (top right) and mass (bottom) distributions at the LHC obtained using the \texttt{MC@NLO/Herwig++} and \texttt{POWHEG} methods (cteq5m/MSbar) and LO (MRST2001LO) for a purely right-handed $W'$ of mass $2\tev$ and width $72\gev$. The invariant mass range and $W'$ mass and widths are identical to those in the previous figures.}
\label{fig:wypzm}
\end{figure}

The subsequent figures in this section represent simulations made for the forthcoming CERN LHC running at 14$\tev$ proton-proton centre of mass energy, expected to run in the second half of 2009.
figure~\ref{fig:scale} shows the variation of the NLO cross section for a 1 TeV left-handed $W'$ in the invariant mass range [400,5000] GeV with factorisation scale, $\mu_F$, for fixed renormalisation scale using the MSbar scheme. The LO variation with PDF scale is also shown in an equivalent range.  The values have been normalised to the cross sections at the default scales $\mu_{0}=\sqrt{k_T^2 + Q^2}$ (default NLO) and $\mu_{0}=Q$ (default LO). In the NLO case the renormalization scale was held fixed at $M_{W'}$. The NLO cross section calculation appears to be slightly more stable than the LO calculation. The $K$-factor at $\mu_{0}$ was found to be $K=1.288$ and the LO cross section at $\mu_{0}=Q$ was found to be $\sigma_{LO}=(2.99\pm0.07)$pb.
Figures~\ref{fig:mt_lh} and~\ref{fig:mt_rh} show the transverse mass distributions at LO and NLO for a $W'$ at masses and widths of $(1\tev,36\gev)$ and $(2\tev,72\gev)$, for purely left-handed ($h_{W'} = 1$) and purely right-handed ($h_{W'} = -1$) couplings to fermions respectively. Figures~\ref{fig:wpt_lh} and~\ref{fig:wpt_rh} show the corresponding $W/W'$ transverse momentum distributions. In this case the LO distribution cuts off at the $W'$ mass. The effect is less visible for higher $W'$ masses. Figure~\ref{fig:wypzm} shows a comparison, for the different methods, of the $W/W'$ rapidity, longitudinal momentum and mass distributions for a right-handed $W'$ of mass $2\tev$ and width $72\gev$ at the LHC. 

Finally, figure~\ref{fig:mt_mcnlo_LR} shows a comparison between the left- and right-handed transverse mass ($M_T$) distributions at NLO, using the \texttt{MC@NLO/Herwig++} method. The importance of the interference between the SM $W$ and the $W'$ can be clearly observed: the differential cross-section in the region below $M_T = M_{W'}$ in the purely left-handed case is reduced in comparison to the purely right-handed case. For transverse mass greater than the on-shell mass of the $W'$, the interference term becomes positive for the left-handed case, although the effect is not significant. The SM contribution, in the absence of a $W'$ boson, is given for comparison in both figures. It can be observed that in the right-handed case the contribution of the $W'$ is simply additive to the SM contribution.
\section{Extraction of limits}\label{sec:limits}
\begin{figure}[htb]
  \vspace{0.5cm}
    \includegraphics[scale=0.8]{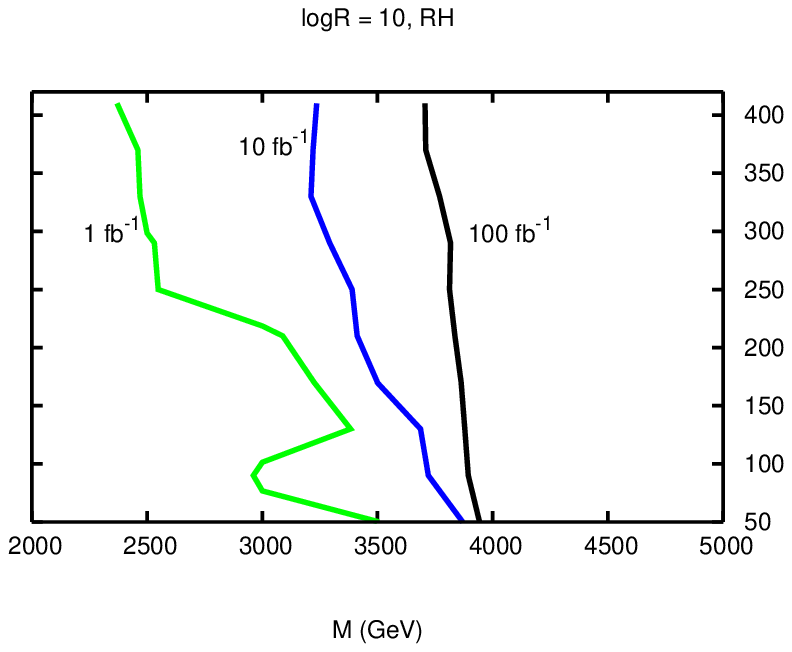}
    \includegraphics[scale=0.8]{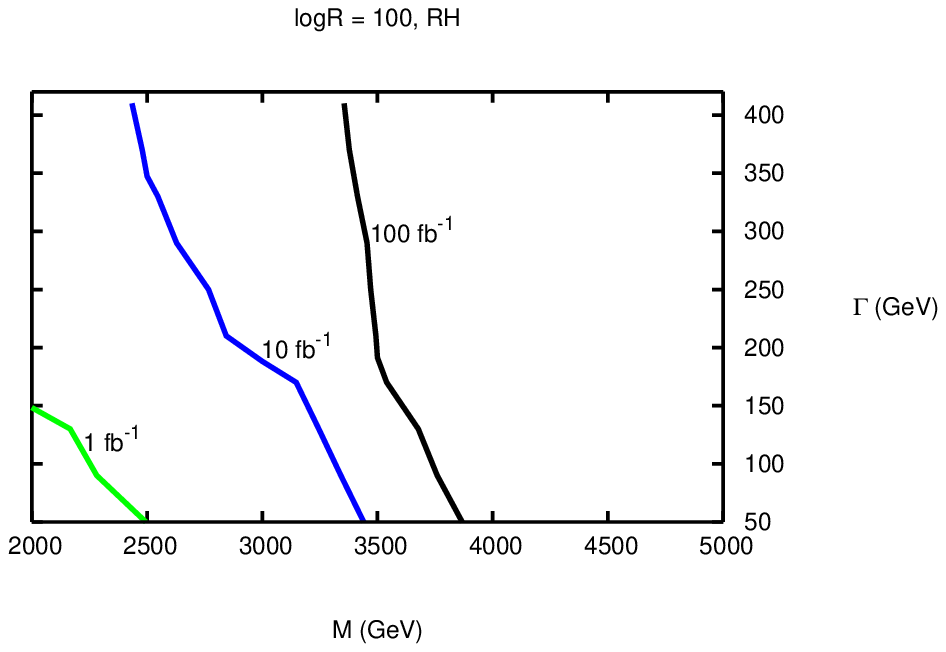}
\caption{The detection reach at the LHC for $\log R = 10$ (left) and $\log R = 100$ (right) at different integrated luminosities for the right-handed case. The colour scheme is: green, blue, black corresponding to the luminosities 1, 10, 100 fb$^{-1}$.}
\label{fig:ex_rh_lhc}
\end{figure}

\begin{figure}[htb]
   \includegraphics[scale=0.8]{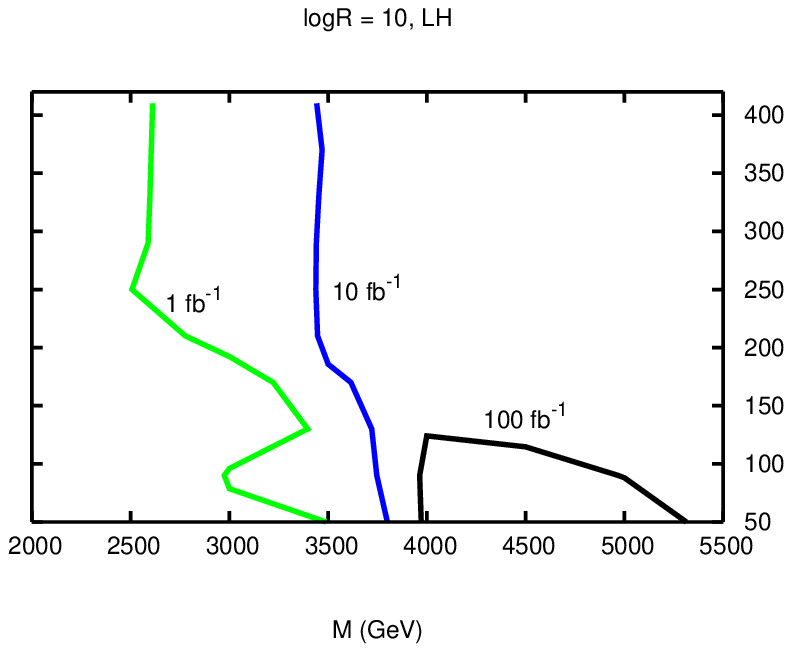}
    \includegraphics[scale=0.8]{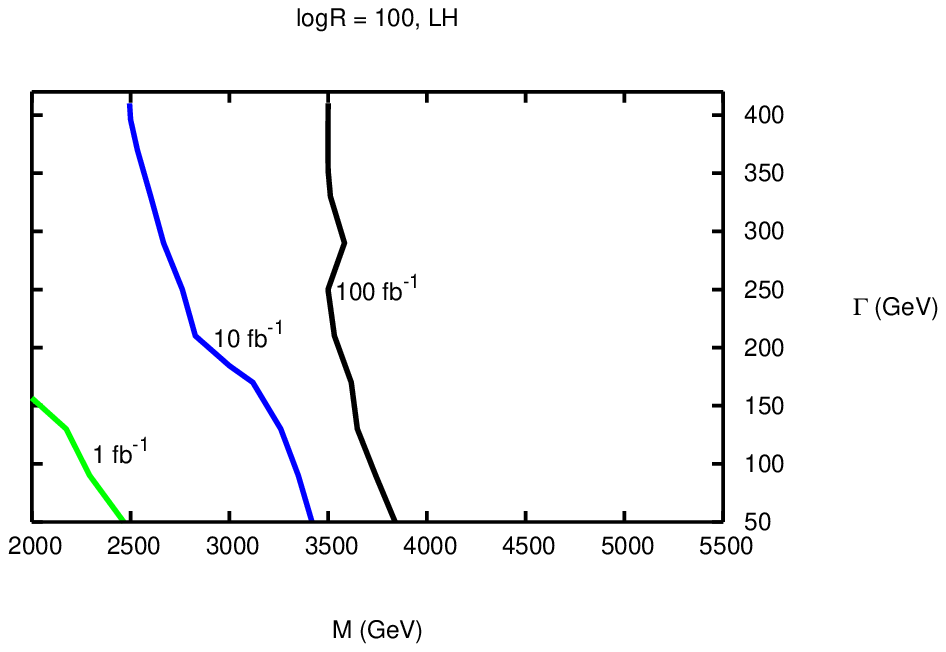}
\caption{The detection reach at the LHC for $\log R = 10$ (left) and $\log R = 100$ (right) at different integrated luminosities for the left-handed case. The colour scheme is identical to the previous figure. In the $\log R = 10$ and 100 fb$^{-1}$ case all points \textit{below} the contour have $\log R < 10$. }
\label{fig:ex_lh_lhc}
\end{figure}

\begin{figure}[htb]
    \includegraphics[scale=0.8]{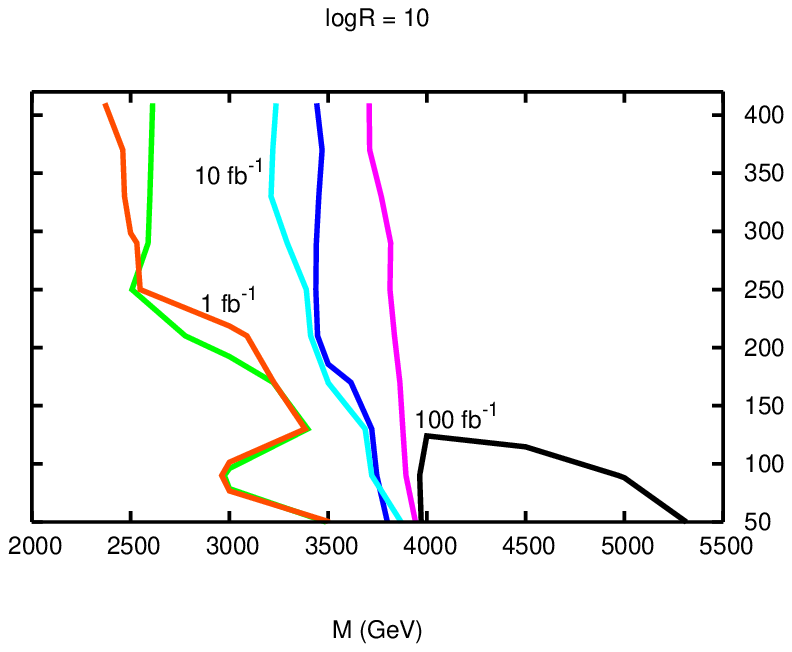}
    \includegraphics[scale=0.8]{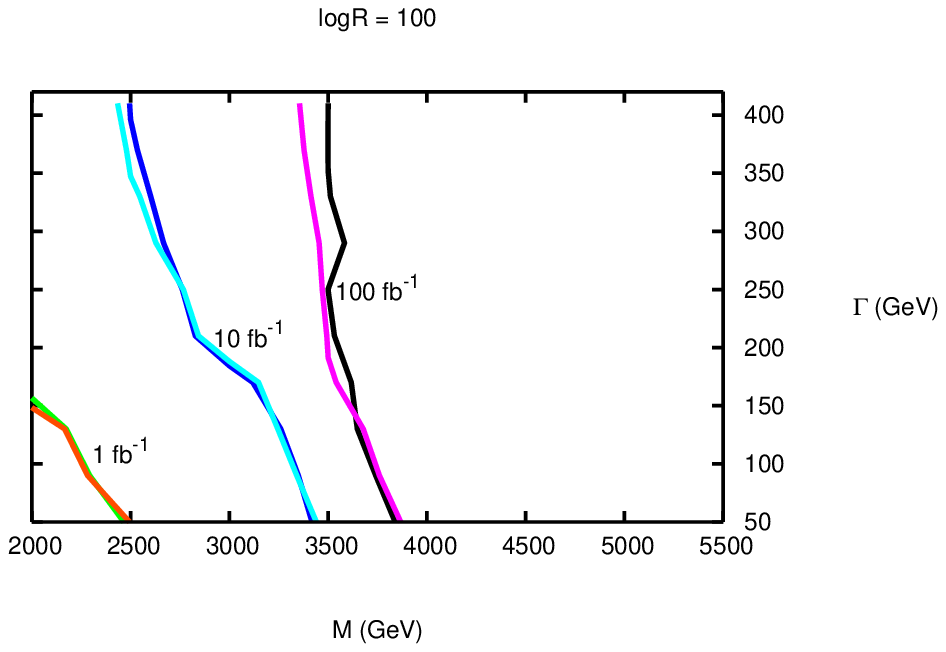}
\caption{The detection reach at the LHC for $\log R = 10$ (left) and $\log R = 100$ (right) at different integrated luminosities for the left- and right-handed cases. The colour scheme is for 1, 10, 100 fb$^{-1}$ is: left-handed: green, blue, black and right-handed: orange, light blue, pink.}
\label{fig:ex_lhc_both}
\end{figure}

In Appendix~\ref{sec:discrimination} we provide a general method for discriminating between two models given a set of data. Here we apply this method to extract observation limits on the $W'$ mass and width at LO. A stand-alone program was written to calculate the quantity $R$ at matrix element-level, given by eq.~\ref{eq:Rpoissonexplog}:
\begin{equation}\label{eq:Rpoissonexplog2}
R \frac{p(S)}{p(T)} = \exp{\left(\sum_{i=0}^N \log\left( \frac{ p(M_{T,i}|T) } { p(M_{T,i} | S) } \right) \right)} \times \left( \frac{\bar{N_T}}{\bar{N_S}}\right)^N e^{-(\bar{N_T} - \bar{N_S} ) }\nonumber
\end{equation}
The `true' underlying theory, called T, was taken to contain a $W'$ at a pre-defined mass and theory S was taken to be the SM. Some comments are appropriate:
\begin{itemize}
\item Although the total $W'$ width was being varied, the decay width to fermions was always taken to be $\Gamma_{W'\rightarrow ff'} = (4\Gamma_{W}/ 3M_{W}) M_{W'} \approx 36 \times M_{W'}$ ($M_{W'}$ in GeV).
\item In the experimental case the $W'$ mass would be unknown and maximum likelihood methods should be used to fit the parameters if significant deviation from the SM is found.
\item The $R$ parameter can become very large if a small number of unlikely events occur, which favour one theory over the other. This is not realistic experimentally since unlikely events could arise from background or detector effects. To take into account these effects, one has to introduce nuisance parameters whose behaviour, at this level of analysis, has to be chosen arbitrarily. In the present analysis we avoid the introduction of such arbitrary parameters.
\item The exclusion curves were drawn for specific data distributions and fluctuations are expected. In other words, the plots given correspond to a specific experimental data set. 
\item The ratio of the prior probability distributions, $p(S)/p(T)$, was taken to be equal to unity throughout this analysis: i.e. we assume both models are equally likely prior to the `experiment'. 
\item A rapidity cut on the leptons corresponding to $y_{cut} = 2.5$ for the LHC and $y_{cut} = 1.3$ for the Tevatron was applied to take into account the acceptance regions of the detectors.
\item The distributions $p(M_T|S)$ and $p(M_T|T)$ were calculated using the Monte Carlo event generator itself at higher statistics ($\sim 10^5$) than the required number of events to reduce the required computer time. The sum over $i$ in eq.~\ref{eq:Rpoissonexplog} was taken over the \textit{bins} of these distributions and not individual events. 
\end{itemize}
The limits were drawn on a width-mass plane as $\log R = C$ ($C$ is a constant) exclusion curves. The variable $R$ can be interpreted as a probability ratio and an exclusion curve $\log R = C$ can be interpreted as the limit where the existence of a $W'$ is excluded with certainty $1 - e^{-C}$. For example if $C = 10$, then the exclusion curve represents the $\sim99.9996\%$ confidence level. The LO exclusion curves can be seen, for different integrated luminosities at the LHC (14 TeV), in figure~\ref{fig:ex_rh_lhc} for a right-handed $W'$ and figure~\ref{fig:ex_lh_lhc} for a left-handed $W'$. The curves correspond to a single data sample at each $(M_{W'}, \Gamma_{W'})$ point, and therefore there are large statistical fluctuations, particularly in the low-luminosity curves. 
A comparison between the curves for a left- and right-handed $W'$ is shown in figure~\ref{fig:ex_lhc_both}. It can be observed that a left-handed $W'$ has a slightly higher detection reach, especially at higher widths. By examining figures~\ref{fig:ex_rh_lhc} and~\ref{fig:ex_lh_lhc}, we can deduce that the maximum detection reach at the LHC, for example using an integrated luminosity of 100 fb$^{-1}$, for a $W'$ decaying primarily to fermions ($\Gamma_{W'} \approx \Gamma_{W'\rightarrow ff'}$), is $\sim 4 \tev$. 
We also show the expected limit at the Tevatron (1.96 TeV) in figure~\ref{fig:ex_tvt_both} with an integrated luminosity of 2 fb$^{-1}$, both at leading and next-to-leading (see below) orders. Note that the current experimental limit on the $W'$ mass is $1\tev$, extracted from a sample of 1 fb$^{-1}$ of data from the D0 experiment~\cite{enu2}. When the $W'$ is only allowed to decay to fermions, i.e. has width $\Gamma_{W'} \approx 36 \gev$, the predicted detection limit for $\log R \sim 10$ is $M_{W'} \approx 1.1 \tev$. This is slightly better than the current Tevatron limit, but is expected to be reduced by experimental effects. Since the available centre-of-mass energy at the Tevatron is $1.96\tev$, we expect the saturation of the detection reach to come at about $ M_{W'} \sim 1 \tev$ without interference and slightly higher in the left-handed case when interference effects are included. The Tevatron NLO case does not exhibit any substantial difference from the LO case.
\begin{figure}[htb]
  \vspace{0.5cm}
    \includegraphics[scale=0.8]{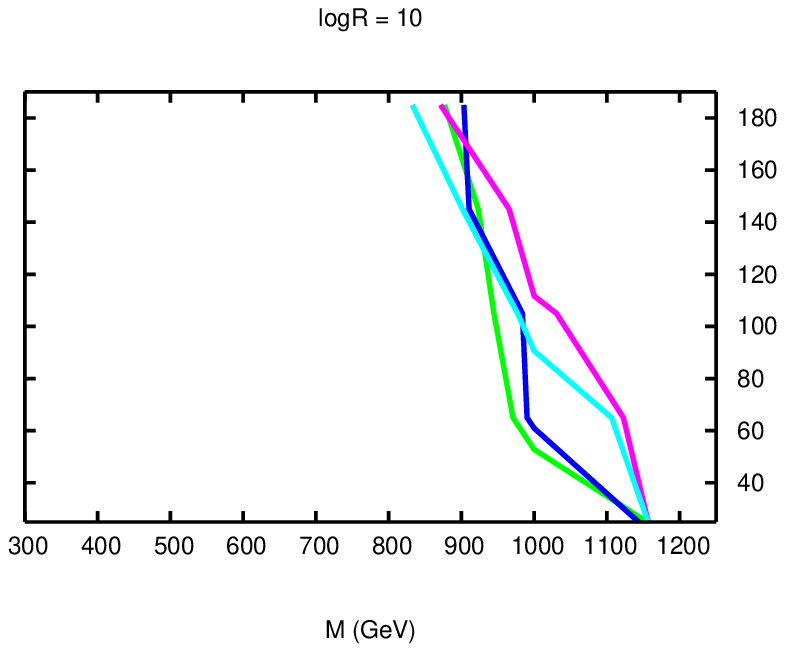}
    \includegraphics[scale=0.8]{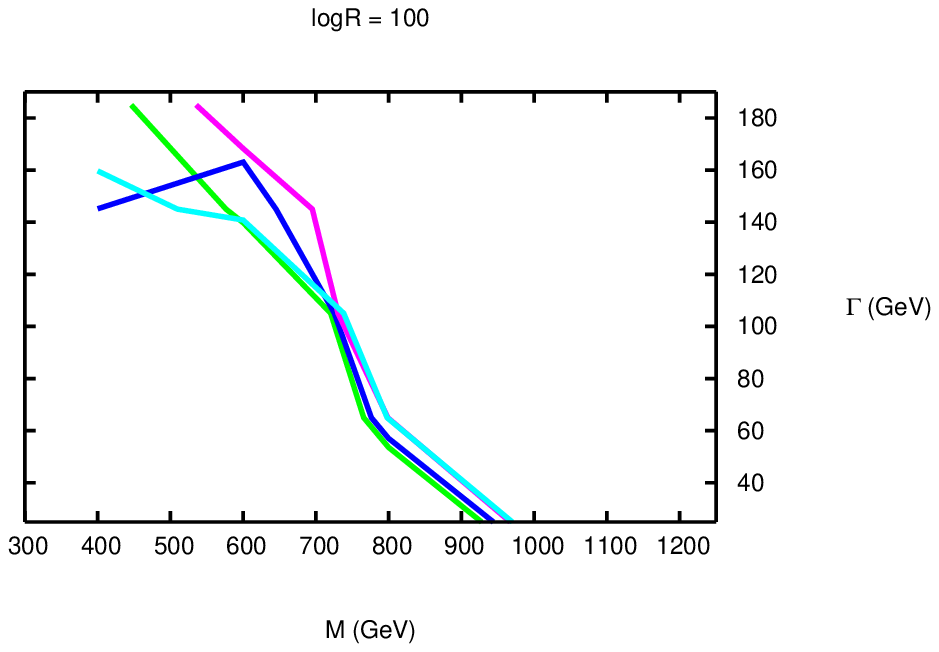}
\caption{The detection reach at the Tevatron for $\log R = 10$ (left) and $\log R = 100$ (right) at 2 fb$^{-1}$, for the left- and right-handed cases, at LO and NLO. The colour scheme is for right-handed and left-handed correspondingly, at LO: green, blue and NLO: light blue, pink.}
\label{fig:ex_tvt_both}
\end{figure}

\begin{figure}[htb]
  \vspace{0.5cm}
    \includegraphics[scale=0.8]{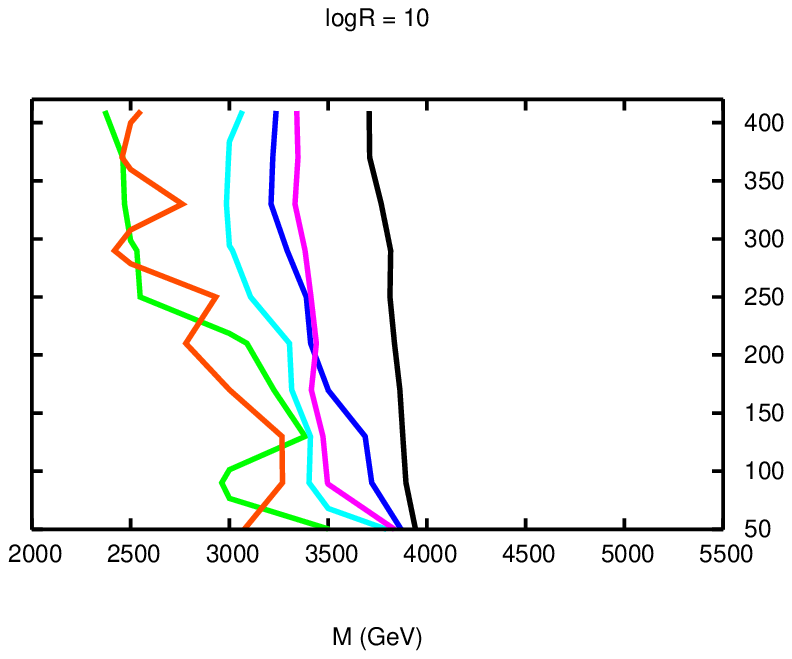}
    \includegraphics[scale=0.8]{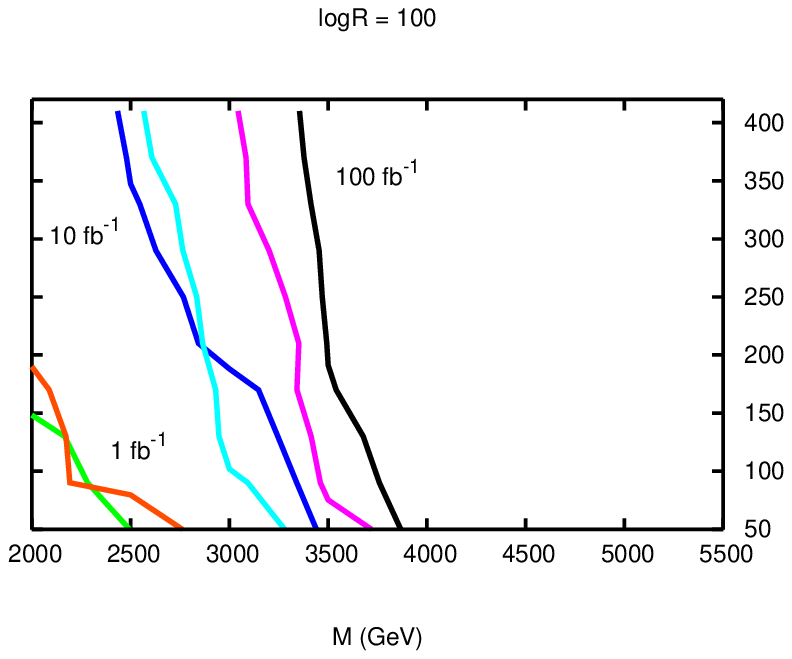}
\caption{The detection reach at the LHC for $\log R = 10$ (left) and $\log R = 100$ (right) at different integrated luminosities for the right-handed case compared at LO and NLO. The colour scheme for 1, 10, 100 fb$^{-1}$ is: LO: green, blue, black and NLO: orange, light blue, pink.}
\label{fig:ex_rh_lhclnlo}
\end{figure}

\begin{figure}[htb]
  \vspace{0.5cm}
  \includegraphics[scale=0.8]{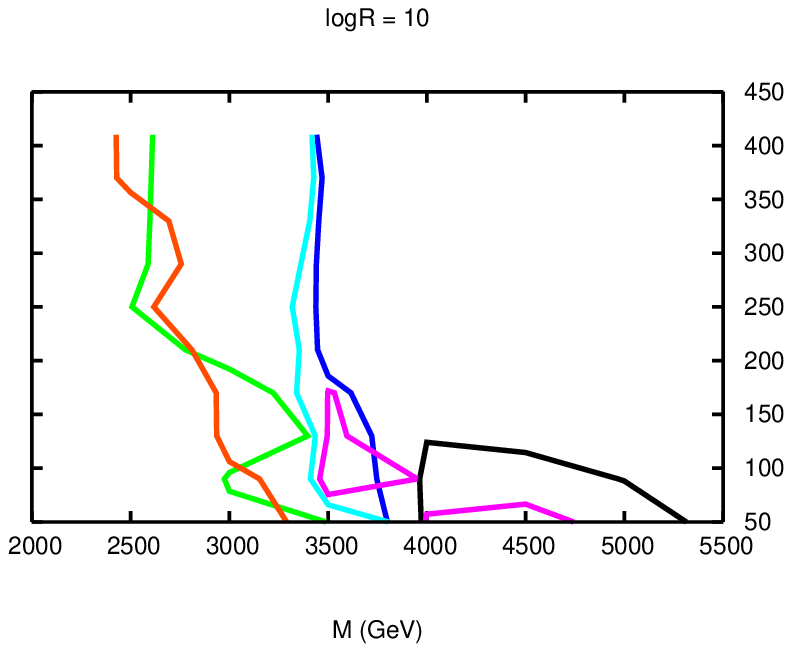}
  \includegraphics[scale=0.8]{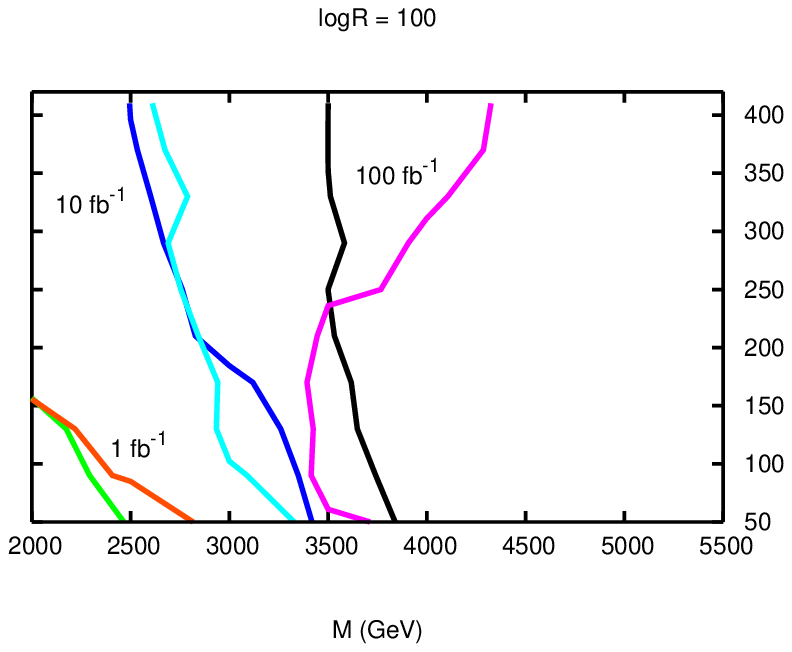}
\caption{The detection reach at the LHC for $\log R = 10$ (left) and $\log R = 100$ (right) at different integrated luminosities for the left-handed case compared at LO and NLO. The colour scheme is identical to the previous figure.}
\label{fig:ex_lh_lhc_lnlo}
\end{figure}

We have performed an equivalent analysis using the NLO method \texttt{POWHEG} at matrix element level to improve computational time. Working at matrix element level with the \texttt{POWHEG} method is justified since the transverse mass distribution is not significantly altered after shower and hadronisation and no difficulties arise due to negative-weighted events, as would be the case in the \texttt{MC@NLO} case. The comments given at the beginning of the section for the LO analysis also apply to the NLO analysis. The results are shown in figures~\ref{fig:ex_rh_lhclnlo} and~\ref{fig:ex_lh_lhc_lnlo} in comparison to the LO results. In the right-handed chirality scenario, NLO implies a lower detection reach than indicated at LO. The situation is more complicated in the left-handed case where the NLO case implies a slightly higher reach for larger widths.  

\begin{figure}[htb]
  \vspace{0.5cm}
  \includegraphics[scale=0.8]{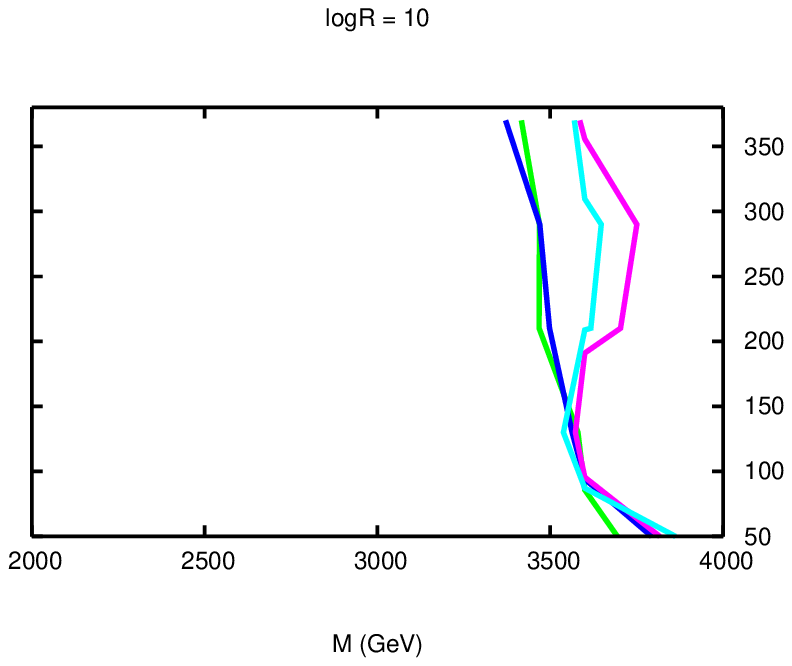}
  \includegraphics[scale=0.8]{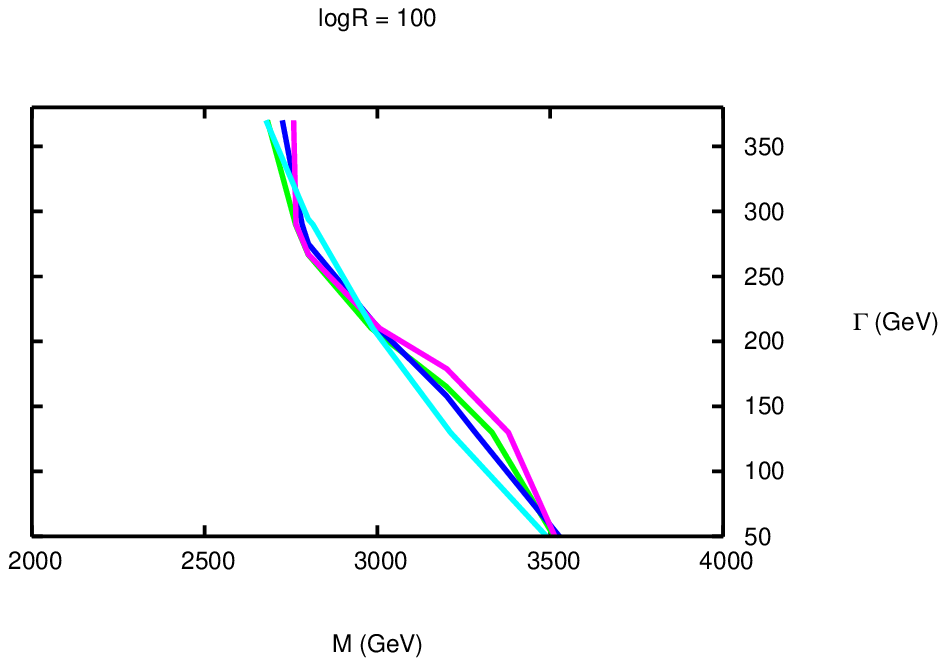}
\caption{The NLO detection reach at the LHC for $\log R = 10$ (left) and $\log R = 100$ (right) for an integrated luminosity of 10 fb$^{-1}$ at different factorisation scale $\mu_F$. The colour scheme for $\mu_F = 0.5\mu_0$, $\mu_0$, $2 \mu_0$ and $4 \mu_0$ is: green, blue, pink and light blue.}
\label{fig:ex_facscale}
\end{figure}

To investigate the dependence of the NLO results on the factorisation scale $\mu_F$ we have reproduced the $\log R$ contours for the right-handed $W'$ LHC case with an integrated luminosity of 10 fb$^{-1}$ at different values of $\mu_F$ while keeping the normalisation scale fixed, using the MSbar scheme. The results are shown in figure~\ref{fig:ex_facscale}. The curves show that the factorisation scale does not affect the detection reach substantially, for example only shifting the $\log R = 10$ contour at a width of $\Gamma_{W'} \sim 200\gev$ from $M_{W'} \sim 3500 \gev$ to $M_{W'} \sim 3750\gev$ going from $\mu_F = 0.5 \mu_0$ to $\mu_F = 4 \mu_0$. 
\section{Conclusions}\label{sec:conclusion} 
We have presented a Monte Carlo implementation of the Drell-Yan production of new charged heavy vector bosons. We have considered the interference effects with the Standard Model $W$ boson, allowing arbitrary chiral couplings to the leptons and quarks. Moreover, the implementation is correct up to next-to-leading order in QCD, via the \texttt{MC@NLO/Herwig++} and \texttt{POWHEG} methods. We have presented a sample of results at both leading and next-to-leading orders. As expected, the LO and NLO boson transverse momentum distributions were found to differ significantly, the NLO extending to higher $p_T$. The dilepton transverse mass, invariant mass, rapidity and $z$-momentum distributions were found not to be significantly altered by the NLO treatment. The total cross section was found to increase in the NLO case by a factor of $\sim 1.3$ in the region of interest. 

Subsequently we applied a theoretical discrimination method to the $W'$ reference model to obtain mass-width observation curves for left- and right-handed chiralities of the $W'$ both at LO and NLO (\texttt{POWHEG}). The NLO curves were shown not to vary significantly with factorisation scale. The event generator used throughout this analysis, \texttt{Wpnlo}, is fully customisable and publicly available~\cite{webpage}. 
\section*{Acknowledgements}
We are grateful to the other members of the {\tt Herwig++} collaboration for developing the program that underlies the present work and for helpful comments. We are also grateful to the Cambridge Supersymmetry Working Group for productive comments. We are particularly grateful to Bryan Webber, Mike Seymour and Marco Sampaio for constructive comments and discussions throughout. This work was supported by the UK Science and Technology Facilities Council, formerly the Particle Physics and Astronomy Research Council, and the European Union Marie Curie Research Training Network MCnet, under contract MRTN-CT-2006-035606.
\appendix
\section{Model discrimination}\label{sec:discrimination}
The search for new physics is essentially a task of discriminating between two models: one with new physics, the other without. The actual task of finding any new physics though is laborious: one has to understand the detector well enough and has to be able to obtain enough statistical significance to say with certainty that something new has been discovered. Here we adopt a rather theoretical approach: we describe a purely statistical method for discriminating between models~\cite{Athanasiou:2006ef, jennie}. This will  essentially yield an upper bound on the detection reach of a heavy particle: detector effects and backgrounds will result in a reduced detection limit. It is useful, however, to be aware of the theoretical possibilities for discovery.
\subsection{Likelihood ratios of probability density functions}\label{sec:likelihood}
Consider $N$ data points, of a mass variable measurement, $\{m_i\}$. Based on these data points, a theoretical model T is $R$ times more likely than another theoretical model S, if,
\begin{equation}\label{eq:Rdef}
R = \frac{ p(T|\{m_i\}) }{ p(S|\{m_i\}) }
\end{equation}
where $p(X|\{m_i\})$ is the probability of model X being true given the data set $\{m_i\}$. We may use Bayes' Theorem to rewrite $R$ as 
\begin{equation}\label{eq:RBayes}
R = \frac{ p(\{m_i\}|T) p(T) }{ p(\{m_i\}|S) p(S) }
\end{equation}
where $p(T)$ and $p(S)$ are the probabilities that S and T are true respectively, usually called prior probabilities since they represent previous knowledge on the theories. We assume that these quantities are equal: there is no strong evidence for either model. We may simplify eq.~\ref{eq:RBayes} further:
\begin{eqnarray}\label{eq:Rexplog}
R \frac{p(S)}{p(T)} = \frac { \Pi_{i=0}^{N} p(m_i|T) } { \Pi_{j=0}^{N} p(m_j|S) } = \Pi_{i=0}^N \frac{ p(m_i|T) } { p(m_i | S) } \nonumber\\
\Rightarrow R \frac{p(S)}{p(T)} = \exp \sum_{i=0}^N \log \left( \frac{ p(m_i|T) } { p(m_i | S) } \right) 
\end{eqnarray}
where we have assumed that events in the data set $\{m_i\}$ are independent. eq.~\ref{eq:Rexplog} is a discrete version of the Kullback-Leibler distance~\cite{kldistance}, a useful quantity for comparing the relative likelihood of two theories according to a data sample. However, it is important to note that the distributions $p(m_i|T)$ and $p(m_i|S)$ are normalized to unity. This means that any difference in the number of events predicted by the two theories will not be taken into account. This will obviously underestimate the significance of a difference in number of events, for example a substantial excess of events in an invariant mass peak. We describe a method which takes this factor into account in the next section.
\subsection{Poisson likelihood ratios}\label{sec:poisson}
In this modification to the method described in the previous section, we simply multiply the variable $R$ defined in eq.~\ref{eq:Rdef} by a ratio of Poisson distributions for the total number of events:
\begin{equation}\label{eq:Rpoissondef}
R = \frac{ p(T|\{m_i\}) }{ p(S|\{m_i\}) } \left( \frac{\bar{N_T}}{\bar{N_S}}\right)^N e^{-(\bar{N_T} - \bar{N_S} ) }
\end{equation}
where $\bar{N_X} = \sigma_X . L$ is the expectation value of the number of events according to theory X, given by the product of the cross-section, $\sigma_X$, and the integrated luminosity, $L$. This expression can be manipulated in a similar manner to eq.~\ref{eq:Rexplog} to obtain:
\begin{equation}\label{eq:Rpoissonexplog}
R \frac{p(S)}{p(T)} = \exp{\left(\sum_{i=0}^N \log\left( \frac{ p(m_i|T) } { p(m_i | S) } \right) \right)} \times \left( \frac{\bar{N_T}}{\bar{N_S}}\right)^N e^{-(\bar{N_T} - \bar{N_S} ) }
\end{equation}
For convenience we may define the `shape' and `Poisson' factors respectively:
\begin{eqnarray}
R_S = \exp \sum_{i=0}^N \log\left( \frac{ p(m_i|T) } { p(m_i | S) } \right)\nonumber \\
R_P = \left( \frac{\bar{N_T}}{\bar{N_S}}\right)^N e^{-(\bar{N_T} - \bar{N_S} ) }
\end{eqnarray}
This method takes into account the difference in the total number of events expected according to each theory at the given integrated luminosity. This is accomplished by re-weighting the `shape' factor $R_S$ by a factor $R_P$ which gives the ratio of probabilities to obtain the observed number of events.
\subsubsection{Application to a toy model}
\begin{figure}[htb]
  \begin{center}
  \input{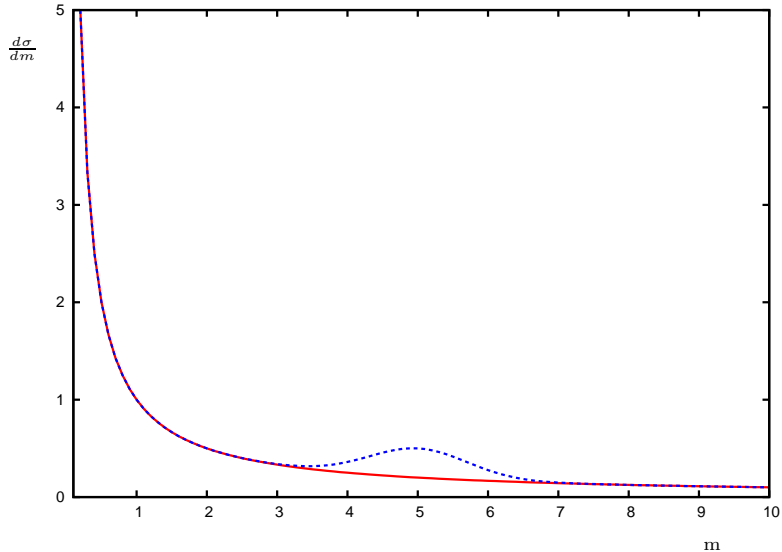}
  \end{center}
\caption{The differential cross-sections $\frac{ \mathrm{d}\sigma } { \mathrm{d} m }$ according to two `toy' theories T and S are shown. Theory T possesses a Gaussian `bump', at $m = 5$ whereas S is just a falling distribution, $1/m$. $m$ is in arbitrary mass units and $\sigma$ in equivalent inverse area squared units.}
\label{fig:TS}
\end{figure}
Before applying the method to the full $W'$ model, it is instructive to present its application to a simple model involving two analytical `toy' distributions. Events for the two distributions have been generated by the general Monte Carlo event generation method. The `differential cross-sections' for the two `theories' T and S with respect to a variable $m$ (in arbitrary units) are given by (defined in the range $[0.1,10]$):
\begin{eqnarray}\label{eq:TSxsection}
\frac{ \mathrm{d}\sigma_T } { \mathrm{d} m } &=& \frac{1}{m} + 0.3 e^{-(m-5)^2}\\ 
\frac{ \mathrm{d}\sigma_S } { \mathrm{d} m } &=& \frac{1}{m} 
\end{eqnarray}
Theory T has a Gaussian peak at $m=5$ on top of a background falling as $\sim 1/m$ and theory S falls as $\sim 1/m$. The situation is shown in figure~\ref{fig:TS}. This is qualitatively similar to the SM tail (theory S) and the SM plus a heavy particle (theory T). The `cross-sections' in the range $m = [0.1,10]$ were found to be $\sigma_T = 5.14$ and $\sigma_S = 4.60$, in arbitrary area units. Assuming an integrated `luminosity' of $L=30$ (equivalent arbitrary inverse area units), we have an expected number of events $\bar{N_T} = 154$ and $\bar{N_S} = 138$. We assume that theory T is the correct underlying theory: events are actually distributed according to it. The result for the variable $R$ if theory T was `true' was found to be $R = 62$. This implies that theory T is 62 times more likely than theory S given the specific data set. If, however, the underlying theory is chosen to be S, then $R = 0.23$. Note that in the case that theory T is `true', it is easier to exclude theory S than to exclude theory T in the case that theory S is `true'.

\clearpage
\bibliography{wppaper}
\bibliographystyle{utphys}
\end{document}